\documentclass[pra,twocolumn,showpacs,superscriptaddress,floatfix]{revtex4-1}
\usepackage[caption=false]{subfig}
\usepackage{graphicx}
\graphicspath{ {} }
\usepackage{amsmath,graphicx}
\usepackage{epsfig}
\usepackage{amsmath,amssymb,mathrsfs,esint}
\usepackage{gensymb}
\usepackage[bookmarks=true,
   colorlinks=true,
   linkcolor=blue,
   urlcolor=blue,
   citecolor=blue,
   bookmarks=true,
   hyperindex=true
]{hyperref}
\usepackage{natbib}
\usepackage{relsize}
\usepackage{float}
\usepackage{dsfont}

\newcommand{\be}{\begin{equation}}
\newcommand{\ee}{\end{equation}}
\newcommand{\ket}[1]{\left| #1 \right>}

\begin{document}

\title{Two-body relaxation of spin-polarized fermions in reduced dimensionalities near a $p$-wave Feshbach resonance}

\author{D.\,V. Kurlov}

\affiliation{Van der Waals-Zeeman Institute, Institute of Physics, University of Amsterdam, Science Park 904, 1098 XH Amsterdam, The Netherlands}
\affiliation{LPTMS, CNRS, Univ. Paris-Sud, Universit\'e Paris-Saclay, 91405 Orsay, France}

\author{G.\,V. Shlyapnikov}

\affiliation{Van der Waals-Zeeman Institute, Institute of Physics, University of Amsterdam, Science Park 904, 1098 XH Amsterdam, The Netherlands}
\affiliation{LPTMS, CNRS, Univ. Paris-Sud, Universit\'e Paris-Saclay, 91405 Orsay, France}
\affiliation{SPEC, CEA, CNRS, Univ. Paris-Saclay, CEA Saclay, Gif sur Yvette 91191, France}
\affiliation{Russian Quantum Center, Novaya Street 100, Skolkovo, Moscow Region 143025, Russia}
\affiliation{State Key Laboratory of Magnetic Resonance and Atomic and Molecular Physics, Wuhan Institute of Physics and Mathematics, Chinese Academy of Sciences, Wuhan 430071, China}

\begin{abstract}
We study inelastic two-body relaxation in a spin-polarized ultracold Fermi gas in the presence of a $p$-wave Feshbach resonance. It is shown that in reduced dimensionalities, especially in the quasi-one-dimensional case, the enhancement of the inelastic rate constant on approach to the resonance is strongly suppressed compared to three dimensions. This may open promising paths for obtaining novel many-body states.
\end{abstract}

\maketitle

\section{Introduction}

Recent progress in the field of ultracold atomic quantum gases opened fascinating prospects to explore novel quantum phases in the systems of degenerate fermions with $p$-wave interactions, for instance two-dimensional (2D) unconventional superfluidity \cite{Gurarie2007}, non-Abelian Majorana modes \cite{Nayak2008, Stern2008}, and itinerant ferromagnetism \cite{Sanner2012, Pekker2011, Jiang2016, Yang2016}. Even though the $p$-wave interactions between cold fermions are much weaker than the $s$-wave interactions, Feshbach resonances allow one to tune the strength and the character of the interactions. However, in the vicinity of such resonances various inelastic collisional processes play a crucial role, resulting in a lifetime of the order of milliseconds at common densities. These are three-body recombination and, for fermionic atoms in an excited hyperfine state, two-body relaxation \cite{Regal2003, Ticknor2004, Chevy2005, Gaebler2007, Levinsen2008, JonaLasinio2008}. 

In this paper we show that in the quasi-2D and quasi-1D geometries the enhancement of two-body inelastic relaxation on approach to the $p$-wave Feshbach resonance is suppressed compared to the three-dimensional (3D) case. This effect is mostly related to a much weaker enhancement of the relative wavefunction near the resonance in reduced dimensionalities. We then demonstrate this for the case of $^{40}$K atoms in the $\ket{F, m_F} = \ket{9/2,-7/2}$ state. A number of experiments was dedicated to the study of atomic fermions in the presence of a $p$-wave Feshbach resonance in quasi-2D and quasi-1D geometries \cite{Moritz2005, Gunter2005, Fuchs2008, Dyke2011}. The atom loss rate has been measured in Ref. \cite{Gunter2005}, and already from this experiment one can see that in reduced dimensionalities the enhancement of the losses near the resonance is reduced compared to the 3D case.

\section{2-body inelastic collisions in 3D}

Let us consider two colliding identical fermions in the vicinity of a $p$-wave Feshbach resonance. In the single-channel model the radial wavefunction of their $p$-wave relative motion at distances $r \gg R_e$, where $R_e$ is a characteristic radius of interaction, has the following form \cite{LL}:
\be \label{psi_3D}
	\psi_{\text{3D}}(r) = i \left\{ j_1(kr) + ik f(k) h_1(k r)  \right\},
\ee
where $k$ is the relative momentum, $j_1(kr)$ and $h_1(kr)$ are spherical Bessel and Hankel functions, and $f(k)$ is the $p$-wave scattering amplitude, which is related to the scattering phase shift $\delta(k)$ as $f(k) = 1/\left[ k( \cot{\delta(k)} - i )\right]$. The $p$-wave S-matrix element is given by $S(k) = \exp 2i\delta(k)$. It is convenient to write the wavefunction (\ref{psi_3D}) at $r\to\infty$ as $\psi_{\text{3D}} = (1/ 2ikr)\left\{ \exp{(-i k r)} + S(k) \exp{(i k r)} \right\}$. In the presence of inelastic collisions the intensity of the outgoing wave is reduced in comparison to the incoming wave by a factor of $|S(k)|^2 < 1$, which implies that the phase shift $\delta (k)$ is a complex quantity with a positive imaginary part. For low collisional energies $E=\hbar^2 k^2/m$ we can use the effective range expansion $k^3 \cot\delta(k) = -1/w_1 - \alpha_1 k^2$, where $w_1$ is the scattering volume and $\alpha_1 > 0$ is the effective range. Then, the scattering amplitude becomes
\be \label{scatt_amp_3D}
	f(k) = \frac{-k^2}{1/w_1 + \alpha_1 k^2 +i k^3},
\ee
and in order to describe inelastic collisions in the vicinity of the resonance, we add an imaginary part to the inverse of the scattering volume: $1/w_1 \to 1/w_1 + i/w'_1$, where $w'_1 > 0$ \cite{MM, Balakrishnan1997}. Therefore, the S-matrix element reads
\be
	S(k) = \frac{1/w_1 +\alpha_1 k^2 + i(1/w'_1 - k^3)}{1/w_1 + \alpha_1 k^2 + i(1/w'_1 + k^3)}.
\ee
For the inelastic rate constant $\alpha_{\text{3D}}(k) = v \sigma^{\text{in}}_{\text{3D}}(k)$, where $v = 2\hbar k/m$ is the relative velocity and $\sigma_{\text{3D}}^{\text{in}}(k) = 3\pi \left[1-|S(k)|^2 \right]/k^2$ is the $p$-wave inelastic scattering cross-section \cite{LL}, we obtain:
\be \label{alpha_3D}
	\alpha_{\text{3D}}(k) = \frac{48\pi\hbar}{m w'_1} \frac{k^2}{\left[ 1/w_1 + \alpha_1 k^2 \right]^2 + \left[ 1/w'_1 + k^3 \right]^2},
\ee
where $m$ is the atom mass and an additional factor of 2 is included since we consider collisions of identical particles. Both $w_1$ and $w'_1$ depend on the external magnetic field, and $w_1$ changes from $+\infty$ to $-\infty$ as one crosses the Feshbach resonance. However, the field dependence of $w'_1$ is weak. Setting $w'_1$ to be field independent we are able to accurately reproduce the results of coupled-channel calculations of the inelastic rate constant and the data of the JILA experiment \cite{Regal2003}. 

Due to the spin-dipole interaction between colliding atoms, the resonant magnetic field (at which the $p$-wave scattering volume diverges) is different for orbital angular momentum projections $m_l = 0$ and $m_l = \pm 1$ \cite{Ticknor2004}. For $^{40}$K atoms in the $\left| 9/2, -7/2 \right>$ state the resonance for $m_l = 0$ occurs at 198.8 G, and for $|m_l| = 1$ at 198.3 G. However, apart from the difference in the position of the resonance, the scattering volume $w_1$ for $m_l = 0$ is the same as it is for $|m_l| = 1$. Moreover, the effective range $\alpha_1$ is also practically the same for all $m_l$'s. In order to clearly demonstrate the effect of suppressed enhancement of 2-body losses near the resonance in reduced dimensionalities, in the main text of the paper we omit the doubling of the resonance due to the spin-dipole interaction. Then the $p$-wave Feshbach resonance for $^{40}$K atoms in the $\left| 9/2, -7/2 \right>$ state in 3D occurs at $B \approx 198.6$ G for all orbital angular momentum projections. Consequently, the rate constant also has a single peak. We discuss the effects associated with the spin-dipole induced doubling of the resonance in the Appendix. 

Sufficiently far from resonance, where the dominant term in the denominator of Eq. (\ref{alpha_3D}) is $1/w_1$, the rate constant becomes
\be \label{alpha_3D_off_res}
	\alpha_{\text{3D}}(k) \approx \frac{48\pi\hbar}{m} \frac{w^2_1}{w'_1}k^2.
\ee

At $T = 0$ we average the rate constant over the Fermi step momentum distribution, and in the off-resonant regime Eq. (\ref{alpha_3D_off_res}) yields
\be \label{alpha_3D_zeroT_off_res}
	\left< \alpha_{\text{3D}} \right>_0 \approx \frac{144 \pi}{5} \, \frac{w_1^2}{w'_1} \frac{E^{\text{3D}}_F}{\hbar},
\ee
where $E_F^{\text{3D}} = \hbar^2 k_F^2 /2m$ is the Fermi energy, $k_F = (6\pi^2 n_{\text{3D}})^{1/3}$ is the Fermi momentum for a single-component 3D gas, and $n_{\text{3D}}$ is the 3D density. 

Near the resonance on its negative side ($w_1 < 0$) the largest contribution to the rate constant comes from momenta close to $\tilde{k}_{\text{3D}} = 1/\sqrt{\alpha_1 |w_1|}$. In the near-resonant regime, where $|w_1|( 1/w'_1 + \tilde{k}^3_{\text{3D}} ) \ll 1$ and $\tilde{k}_{\text{3D}} \ll k_F$, the rate constant exhibits a sharp peak, which is slightly shifted with respect to the position of the resonance at zero kinetic energy ($1/w_1 = 0$). The maximum value of the rate constant can be estimated as
\be \label{alpha_3D_zeroT_res}
	\left< \alpha_{\text{3D}} \right>_0 \approx \frac{576 \pi^2 \hbar}{\alpha_1 \, m} \frac{1}{1+w'_1 \tilde{k}^3_{\text{3D}}} \left(\frac{\tilde{k}_{\text{3D}}}{k_F}\right)^3.
\ee 

Using Eq. (\ref{alpha_3D}) we calculate numerically $\left< \alpha_{\text{3D}} \right>_0$ for $^{40}$K atoms in magnetic fields from 195 to 205 G. The results are presented in Fig. \ref{alpha3D_ZeroT} for $E^{\text{3D}}_F = 1$ $\mu$K and $4$ $\mu$K (corresponding to densities $n_{\text{3D}} \approx 3.6 \times 10^{13}$ cm$^{-3}$ and $2.9 \times 10^{14}$ cm$^{-3}$, respectively).  In order to determine $w'_1$ we fit Eq. (\ref{alpha_3D}) to the results of coupled-channel numerical calculations of the relaxation rate for a gas of $^{40}$K atoms in $\ket{9/2, -7/2}$ state at a fixed collisional energy of 1 $\mu$K \cite{BohnPrivate}, using the values of $w_1$ and $\alpha_1$ that have been measured in the JILA experiment \cite{Ticknor2004}. Then we obtain $w'_1 = 0.53 \times 10^{-12}$ cm$^3$. For the scattering volume $w_1$ and the effective range $\alpha_1$ we take the values measured in the JILA experiment \cite{Ticknor2004} for $|m_l| = 1$ and manually shift the position of the resonance from 198.3~G to 198.6 G, so that the effect of the spin-dipole interaction is compensated. The off-resonant expression (\ref{alpha_3D_zeroT_off_res}) shows perfect agreement with the numerical results, and the near-resonant expression (\ref{alpha_3D_zeroT_res}) leads to a slight overestimate. However, Eq. (\ref{alpha_3D_zeroT_res}) correctly captures that in the vicinity of the maximum $\left< \alpha_{\text{3D}} \right>_0 \sim (E_F^{\text{3D}})^{-3/2}$, in contrast to the off-resonant case, where the rate constant behaves as $\left< \alpha_{\text{3D}} \right>_0 \sim E_F^{\text{3D}}$.

For the classical gas ($T \gg E_F^{\text{3D}}$) averaging $\alpha_{\text{3D}}(k)$ over the Boltzmann distribution of atoms, in the off-resonant regime we obtain:
\be \label{alpha_3D_highT_off_res}
	\left< \alpha_{\text{3D}} \right>_T \approx \frac{72 \pi w_1^2}{w'_1} \,\frac{T}{\hbar}.
\ee

On the negative side of the resonance in the near-resonant regime, where $|w_1|( 1/w'_1 + \tilde{k}^3_{\text{3D}} ) \ll 1$ and $\tilde{k}_{\text{3D}} \ll k_T$, with $k_T = \sqrt{m T /\hbar^2}$ being the thermal momentum, the rate constant has a sharp peak slightly shifted with respect to the resonance at zero kinetic energy:
\be \label{alpha_3D_highT_res}
	\left< \alpha_{\text{3D}} \right>_T = \frac{96 \pi^{3/2}\hbar}{\alpha_1 \, m} \, \frac{1}{1+w'_1 \tilde{k}^3_{\text{3D}}} \left(\frac{\tilde{k}_{\text{3D}}}{k_T}\right)^3.
\ee
\linebreak
Direct numerical calculation of $\left< \alpha_{\text{3D}} \right>_T$ using Eq.~(\ref{alpha_3D}) shows a perfect agreement with both off-resonant and near-resonant expressions, as shown in Fig.~\ref{alpha3D_HighT} for $T = 300$~nK and $T = 1$ $\mu$K. 

Thus, we see that the inelastic rate constant has a drastically different temperature (Fermi energy) dependence in the near-resonant regime compared to the off-resonant case. For deep inelastic collisions the energy dependence of the rate constant is completely determined by the wavefunction of the initial state of colliding particles. In order to gain insight into the behavior of the inelastic rate constant we analyze the behavior of the wavefunction of the relative motion of two atoms at distances where $R_e \ll r \ll k^{-1}$. Using Eq. (\ref{psi_3D}) and the expressions for the amplitude $f(k)$ and the phase shift $\delta(k)$ written after this equation, we have:
\be \label{psi_3D_small_distances}
	\psi_{\text{3D}}(r) \approx i \frac{\left(1/w_1 + \alpha_1 k^2\right) k r / 3 - k/r^2}{1/w_1 + \alpha_1 k^2 + i k^3}.
\ee

In the off-resonant regime the terms containing $1/w_1$ are the leading ones in both the numerator and denominator of Eq. (\ref{psi_3D_small_distances}), and $\psi_{\text{3D}}^{\text{off}} \approx i kr /3$. This leads to $\alpha_{\text{3D}}^{\text{off}} (k) \sim k^2$, in agreement with Eq. (\ref{alpha_3D_off_res}). In the off-resonant regime collisions with all momenta in the distribution function contribute to the inelastic rate constant. In contrast, in the near-resonant regime on the negative side of the resonance ($w_1 < 0$) only a small fraction of relative momenta contributes to $\alpha_{\text{3D}}$. These are momenta in a narrow interval $\delta k \sim \tilde{k}_{\text{3D}}^2 / \alpha_1$ around $\tilde{k}_{\text{3D}}$. Accordingly, in the classical gas the fraction of such momenta is $F_{\text{3D}} \sim \tilde{k}_{\text{3D}}^2 \delta k / k_T^3 \sim \tilde{k}_{\text{3D}}^4/(\alpha_1 k_T^3)$. In this near-resonant regime we have $|1/w_1 + \alpha_1 k^2| \sim \tilde{k}^3_{\text{3D}}$. Then, putting the rest of $k$'s equal to $\tilde{k}_{\text{3D}}$ in Eq. (\ref{psi_3D_small_distances}) and taking into account that $\tilde{k}_{\text{3D}} r \ll 1$ at $r$ approaching $R_e$, we see that the relative wavefunction in the near-resonant regime is $\psi_{\text{3D}}^{\text{res}} \approx 1/(\tilde{k}_{\text{3D}} r)^2$. The ratio of the near-resonant inelastic rate constant to the off-resonant one is ${\cal R}_{\text{3D}}$ $\sim (\psi_{\text{3D}}^{\text{res}} / \psi_{\text{3D}}^{\text{off}})^2 F_{\text{3D}}$, where we have to put $r \sim R_e$ in the expressions for the relative wavefunctions. This yields
\be \label{R_3D}
	{\cal R}_{\text{3D}}\equiv \left< \alpha_{\text{3D}}^{\text{res}} \right>_T / \left< \alpha_{\text{3D}}^{\text{off}} \right>_T \sim 1/(k_T R_e)^5.
\ee
This is consistent with equations (\ref{alpha_3D_highT_off_res}) and (\ref{alpha_3D_highT_res}), since $\alpha_1 \sim 1/R_e$ and $w_1$ in the off-resonant regime is $\sim R_e^3$ (and we may omit unity compared to $w'_1 \tilde{k}_{\text{3D}}^3$ in the denominator of (\ref{alpha_3D_highT_res})).  

\begin{widetext}

\begin{figure}
\subfloat{%
  \includegraphics[width=.49\linewidth]{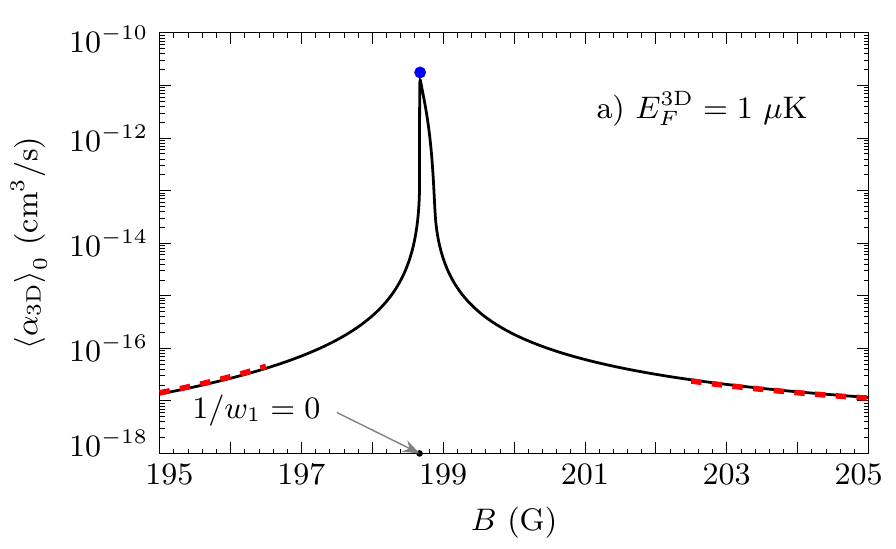}%
}\hfill
\subfloat{%
  \includegraphics[width=.49\linewidth]{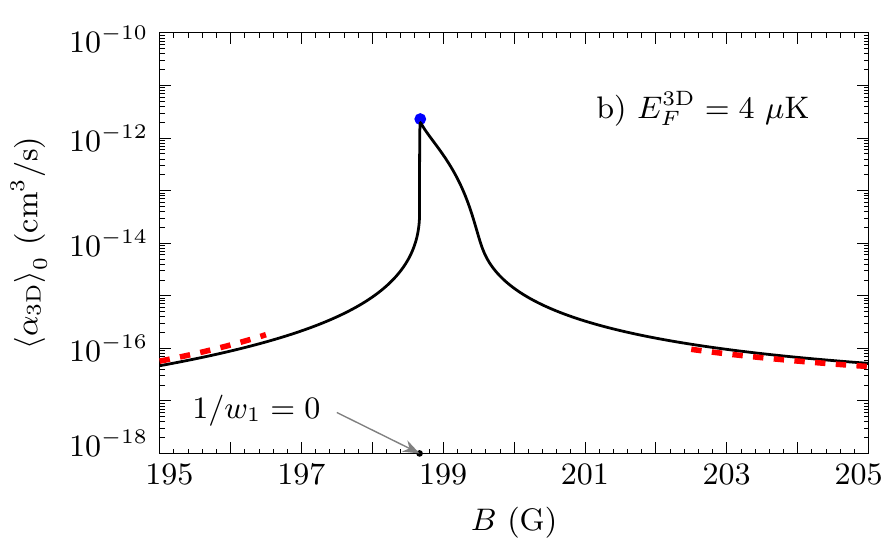}%
}
\caption{Three-dimensional inelastic rate constant $\left<\alpha_{\text{3D}}\right>_0$ for $^{40}$K atoms in the $\ket{9/2,-7/2}$ state at $T = 0$ versus magnetic field $B$ for $E^{\text{3D}}_F = 1$ $\mu$K in a) and $E^{\text{3D}}_F = 4$ $\mu$K in b). Dashed red curves correspond to the off-resonant regime described by Eq. (\ref{alpha_3D_zeroT_off_res}), and the blue point marks the near-resonant peak value given by Eq. (\ref{alpha_3D_zeroT_res}). It is shifted in the direction of higher fields by $7.8$ mG in a) and by $10$ mG in b) with respect to the magnetic field at which $1/w_1 = 0$.}
\label{alpha3D_ZeroT}
\end{figure}

\begin{figure}
\subfloat{%
  \includegraphics[width=.49\linewidth]{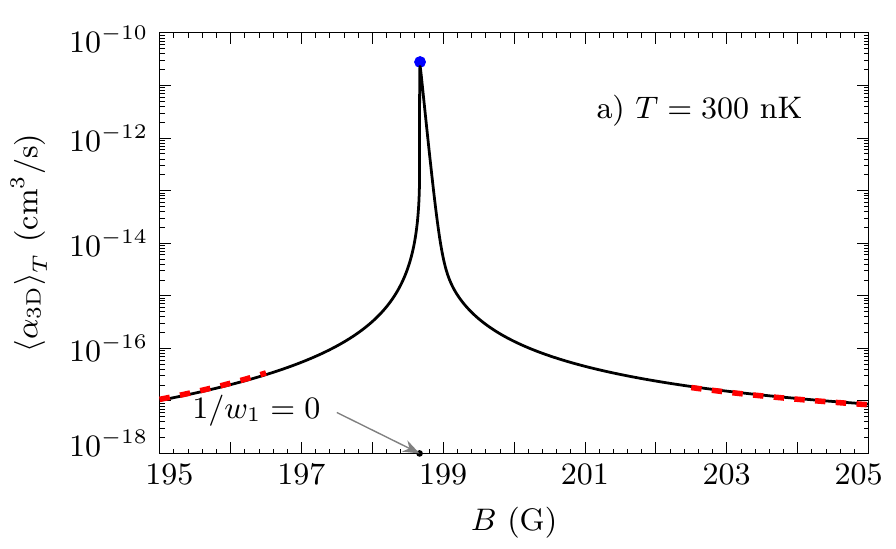}%
}\hfill
\subfloat{%
  \includegraphics[width=.49\linewidth]{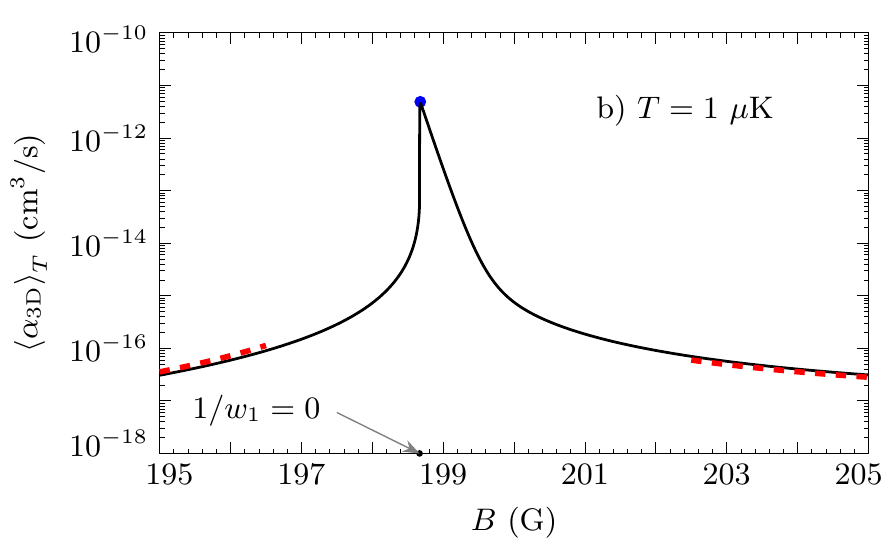}%
}
\caption{Three-dimensional inelastic rate constant $\left<\alpha_{\text{3D}}\right>_T$ for $^{40}$K atoms in the $\ket{9/2,-7/2}$ state versus magnetic field $B$ for $T = 300$ nK in a) and $T = 1$ $\mu$K in b). Dashed red curves correspond to the off-resonant regime described by Eq. (\ref{alpha_3D_highT_off_res}), and the blue point marks the near-resonant peak value according to Eq. (\ref{alpha_3D_highT_res}). It is shifted in the direction of higher fields by $6.7$ mG in a) and by $10$ mG in b) with respect to the magnetic field at which $1/w_1 = 0$.}
\label{alpha3D_HighT}
\end{figure}

\end{widetext}

From Figs. \ref{alpha3D_ZeroT} and \ref{alpha3D_HighT} we see that there is a difference in the asymmetry of the profiles between $\left< \alpha_{\text{3D}} \right>_0$ and $\left< \alpha_{\text{3D}} \right>_T$. This difference can be explained as follows. The resonance takes place for particles with relative momenta close to $\tilde k_{\text{3D}} = 1/\sqrt{\alpha_1 |w_1|}$, which (in $^{40}$K) grows with the magnetic field. At low $E_F$ (at $T = 0$) or low $T$ (in the Boltzmann gas), the number of particles with such momenta is small, but it becomes larger for higher $E_F$ or $T$. Thus the rate constant increases \cite{Ticknor2004}. In a strongly degenerate gas, at a magnetic field corresponding to $\tilde k_{\text{3D}} \sim k_F$, the rate constant $\left< \alpha_{\text{3D}} \right>_0$ rapidly decreases, since due to the Fermi step momentum distribution there are no particles that can experience resonant scattering at higher $B$ fields (Fig. \ref{alpha3D_ZeroT}). In contrast, in a classical gas the high-field tail of $\left< \alpha_{\text{3D}} \right>_T$ decreases towards the off-resonant values more gradually due to the Boltzmann momentum distribution of colliding particles (Fig. \ref{alpha3D_HighT}).

\section{2-body inelastic collisions in 2D}

We now consider inelastic collisions in the two-dimensional case and again omit the doubling of the resonance due to the spin-dipole interaction. The resulting single-peak structure of the relaxation rate constant is realized for the magnetic field perpendicular to the plane of the translational motion. In this case the relative wavefunction at short interparticle distances corresponds to the 3D motion with $|m_l| = 1$. The spin-dipole interaction only shifts the peak of the rate constant, and the discussion of this shift is moved to the Appendix. In the Appendix we also present calculations taking into account the spin-dipole doubling of the resonance and the emerging double-peak structure of the relaxation rate for the magnetic field parallel to the plane of the translational motion.

In the quasi-2D geometry obtained by a tight harmonic confinement in the axial direction ($z$) with frequency $\omega_0$, at in-plane ($x,y$) interatomic separations $\rho$ greatly exceeding the extension of the wavefunction in the axial direction, $l_0=(\hbar/m\omega_0)^{1/2}$, the $p$-wave relative motion is described by the wavefunction
\be \label{psi_2D}
	\psi_{\text{2D}}({\boldsymbol r}) = \varphi_{\text{2D}} ( \rho) e^{i \vartheta}  \frac{1}{(2\pi l_0^2)^{1/4}}\exp\left\{ -\frac{z^2}{4l_0^2} \right\};\,\,\,\,\rho\gg l_0,
\ee
where $\vartheta$ is the scattering angle, and $z$ is the interparticle separation in the axial direction. For remaining in the ultracold limit with respect to the axial motion we will assume below that $l_0\gg R_e$ \cite{Petrov2001}. Then the 2D $p$-wave radial wavefunction $\varphi_{\text{2D}}$ is 
\be \label{phi_2D}
	\varphi_{\text{2D}} (\rho) = i \left\{ J_1( q \rho) - \frac{i}{4} f_{\text{2D}}(q) H_1( q \rho) \right\},
\ee
with $q$ being the 2D relative momentum, and $J_1(q\rho)$ and $H_1(q\rho)$ the Bessel and Hankel functions. The $p$-wave quasi-2D scattering  amplitude $f_{\text{2D}}(q)$ is given by \cite{Pricoupenko2008, Peng2014}
\be \label{scatt_amp_2D}
	f_{\text{2D}}(q) = \frac{4 q^2}{1/A_p + B_p q^2 -(2 q^2 / \pi) \ln l_0 q + i q^2},
\ee
where the 2D scattering parameters are $1/A_p = (4 / 3\sqrt{2\pi} l_0^2) \left[ l_0^3/w_1 + \alpha_1 l_0 / 2 - {\cal C}_1 \right]$ and $B_p = (4/3\sqrt{2\pi}) \left[ l_0 \alpha_1 - {\cal C}_2 \right]$, with numerical constants ${\cal C}_1 \approx 6.5553 \times 10^{-2}$ and ${\cal C}_2 \approx 1.4641 \times 10^{-1}$. The amplitude $f_{\text{2D}}$ is related to the S-matrix element $S_{\text{2D}}$ as $f_{\text{2D}}(q) = 2i(S_{\text{2D}}(q) - 1)$. The 2D (confinement-influenced) resonance occurs at $1/A_p=0$ and is thus shifted with respect to the 3D resonance ($1/w_1=0$). Like in the 3D case, in the presence of inelastic processes we have to replace $1/w_1$ by $1/w_1 + i/w'_1$, which yields
\be
	S_{\text{2D}}(q) = \frac{\mathlarger{\frac{1}{A_p}} + \left(B_p  - \mathlarger{\frac{2 \ln l_0 q}{\pi}}\right) q^2+ i \left(\mathlarger{\frac{1}{A'_p}} - q^2\right)}{\mathlarger{\frac{1}{A_p}} + \left(B_p  - \mathlarger{\frac{2\ln l_0 q}{\pi}} \right) q^2 + i \left(\mathlarger{\frac{1}{A'_p}} + q^2\right)},
\ee
with $A'_p =  3\sqrt{2\pi} w'_1 / 4 l_0 > 0$. Then, writing the wavefunction (\ref{phi_2D}) at $\rho \to \infty$ as $\varphi_{\text{2D}} \approx (1/\sqrt{2\pi i q \rho}) \left\{ \exp{(-i q \rho)} + i S_{\text{2D}}(q) \exp{(i q \rho)} \right\}$ we see that the intensity of the outgoing wave is reduced by a factor of $|S_{\text{2D}}(q)|^2 < 1$ compared to the incoming wave. The 2D inelastic cross-section is defined as $\sigma_{\text{2D}}^{\text{in}} = (2/q)\left(1-|S_{\text{2D}}(q)|^2 \right)$, and for identical particles one has an additional factor of 2. Then, for the inelastic rate constant, $\alpha_{\text{2D}}(q) = (2\hbar q / m) \sigma_{\text{2D}}^{\text{in}}$, we obtain:
 \be \label{alpha_2D}
 \begin{aligned}
 	&\alpha_{\text{2D}}(q) =\frac{32\hbar}{m A'_p} \\
	&\times \frac{q^2}{\left[ \mathlarger{\frac{1}{A_p}} + \left(B_p - \mathlarger{\frac{2 \ln l_0 q}{\pi}}\right) q^2 \right]^2 + \left[ \mathlarger{\frac{1}{A'_p}} + q^2 \right]^2}.
\end{aligned}
\ee
Sufficiently far from the resonance, where the dominant term in the denominator of Eq. (\ref{alpha_2D}) is $1/A_p$, the rate constant becomes
 \be \label{alpha_2D_off_res}
 	\alpha_{\text{2D}}(q) \approx \frac{32\hbar}{m}\frac{A_p^2}{A'_p} q^2 \approx \frac{24\sqrt{2\pi} \hbar}{m l_0} \frac{w_1^2}{w'_1} q^2.
 \ee
At $T = 0$, averaging the off-resonant rate constant (\ref{alpha_2D_off_res}) over the Fermi step momentum distribution we obtain 
\be \label{alpha_2D_zeroT_off_res}
	\left< \alpha_{\text{2D}} \right>_0 \approx \frac{12 \sqrt{2\pi} w_1^2}{l_0 w'_1} \frac{E^{\text{2D}}_F}{\hbar},
\ee
where $E^{\text{2D}}_F = \hbar^2 q_F^2 /2m$ is the 2D Fermi energy, $q_F = \sqrt{4\pi n_{\text{2D}}}$ is the Fermi momentum for a single-component 2D gas, and $n_{\text{2D}}$ is the 2D density.

Near the 2D resonance on its negative side ($A_p < 0$) the largest contribution to the rate constant comes from momenta close to $\tilde{q}_{\text{2D}} = 1/\sqrt{B_p |A_p|}$. Then, in the regime, where $(A'_p)^{-1/2} \ll \tilde{q}_{\text{2D}} \ll q_F$, the rate constant has a sharp peak. The maximum value of the 2D rate constant at $T=0$ can then be estimated as
\be \label{alpha_2D_zeroT_res}
	\left< \alpha_{\text{2D}} \right>_0 \approx \frac{128 \pi \hbar}{A'_p \, B_p \,m}\frac{1}{q_F^2} \approx \frac{128 \pi \hbar}{ \alpha_1 m\, w'_1}\frac{1}{q_F^2},
\ee 
where we took into account that $A'_p B_p = w'_1 (\alpha_1 l_0 - {\cal C}_2)/ l_0 \approx w'_1 \alpha_1$ for typical confinement frequencies $\omega_0$ from $50$ to $150$ kHz. Therefore, the tight harmonic confinement has almost no influence on the maximum value of $\left< \alpha_{\text{2D}} \right>_0$.

The results of direct numerical calculation of $\left< \alpha_{\text{2D}} \right>_0$ using Eq. (\ref{alpha_2D}) for $^{40}$K atoms are presented in Fig. \ref{alpha2D_ZeroT} for the confining frequency $\omega_0 = 120$ kHz and Fermi energies $E^{\text{2D}}_F = 1$ $\mu$K and $4$ $\mu$K (corresponding to densities $n_{\text{2D}} \approx 1.3 \times 10^{9}$ cm$^{-2}$ and $5.2 \times 10^{9}$ cm$^{-2}$, respectively). The off-resonant expression (\ref{alpha_2D_zeroT_off_res}) shows perfect agreement with the numerical results, while the near-resonant expression (\ref{alpha_2D_zeroT_res}) leads to a slight overestimate. However, Eq. (\ref{alpha_2D_zeroT_res}) captures that in the vicinity of the maximum $\left< \alpha_{\text{2D}} \right>_0 \sim 1/E_F^{\text{2D}}$, in contrast to the off-resonant case, where $\left< \alpha_{\text{2D}} \right>_0 \sim E_F^{\text{2D}}$.

At $T \gg E^{\text{2D}}_F$, we average Eq. (\ref{alpha_2D}) over the Boltzmann distribution of atoms. Then, the off-resonant expression for the rate constant follows from Eq. (\ref{alpha_2D_off_res}) and reads
\be \label{alpha_2D_highT_off_res}
	\left< \alpha_{\text{2D}} \right>_T \approx \frac{24 \sqrt{2\pi} w_1^2}{l_0 w'_1} \frac{T}{\hbar}.
\ee

On the negative side of the 2D resonance in the near-resonant regime, where $(A'_p)^{-1/2} \ll \tilde{q}_{\text{2D}} \ll q_T$, with $q_T = \sqrt{m T /\hbar^2}$ being the thermal momentum, the rate constant has a sharp peak slightly shifted from the position of the 2D resonance at zero kinetic energy. The maximum value of the rate constant is given by
\be \label{alpha_2D_highT_res}
	\left< \alpha_{\text{2D}} \right>_T \approx \frac{32 \pi \hbar}{ A'_p B_p \,m}\frac{1}{q_T^2} \approx \frac{32 \pi \hbar}{ \alpha_1 m \, w'_1}\frac{1}{q_T^2}.
\ee
Like in the zero temperature case, we see that the maximum value of $\left< \alpha_{\text{2D}} \right>_T$ is practically independent of the confinement frequency.

Direct numerical calculation of $\left< \alpha_{\text{2D}} \right>_T$ from Eq. (\ref{alpha_2D}) shows perfect agreement with both off-resonant and near-resonant expressions, as displayed in Fig.~\ref{alpha2D_HighT} for the confining frequency $\omega_{0} = 120$ kHz and temperatures $T = 300$~nK and $T = 1$ $\mu$K. 

In order to qualitatively understand the temperature (Fermi energy) dependence of  the inelastic rate constant, we analyze the structure of the initial state wavefunction, which for deep inelastic processes fully determines the energy dependence of $\alpha_{\text{2D}}$. Inelastic collisions occur at interparticle distances $r\lesssim R_e\ll l_0$, where the relative motion of colliding atoms has a three-dimensional character and $\psi_{\text{2D}}$ is different from $\psi_{\text{3D}}$ only by a normalization coefficient. Assuming the inequality $kl_0 \ll 1$, at distances $r$ exceeding $R_e$ sufficiently far from the 3D resonance from Eq. (\ref{psi_3D_small_distances}) we have $\psi_{\text{3D}}(r) \propto \left\{ r - 3w_1/r^2 \right\}$. Then, according to Ref. \cite{Pricoupenko2008}, the 2D wavefunction can be written as
\be \label{psi_2D_small_distances}
	\psi_{\text{2D}}(r) = \frac{if_{\text{2D}}(q)(2\pi l_0^2)^{1/4}}{6\pi w_1q}\left\{ r - 3w_1/r^2 \right\}.
\ee
Far from the 2D resonance ($1/A_p = 0$) the 2D scattering amplitude is $f_{\text{2D}}^{\text{off}}\approx 3\sqrt{2\pi}w_1 q^2 / l_0$, which leads to $\psi_{\text{2D}}^{\text{off}} \sim (q/\sqrt{l_0}) \left\{ r - 3w_1 / r^2 \right\}$ and $\alpha_{\text{2D}}^{\text{off}} \sim q^2 / l_0$, in agreement with Eq. (\ref{alpha_2D_off_res}). In the near-resonant regime on the negative side of the 2D resonance ($A_p < 0$) the main contribution to $\alpha_{\text{2D}}$ is provided by relative momenta in a narrow interval $\delta q \sim \tilde{q}_{\text{2D}} / B_p$ around $\tilde{q}_{\text{2D}}$. In the classical gas the fraction of such momenta is $F_{\text{2D}} \sim \tilde{q}_{\text{2D}} \delta q / q_T^2 \sim \tilde{q}_{\text{2D}}^2 /(B_p q_T^2)$. In this near-resonant regime we have $|1/A_p + B_p q^2| \sim \tilde{q}^2_{\text{2D}}$. Then, we may put the rest of $q$'s equal to $\tilde{q}_{\text{2D}}$ in Eq. (\ref{psi_2D_small_distances}) and use $f_{\text{2D}}^{\text{res}}(q) \approx -4i$ (omitting the logarithmic term in the denominator of $f_{\text{2D}}(q)$). Thus, the 2D wavefunction becomes $\psi_{\text{2D}}^{\text{res}} \sim (\sqrt{l_0} / \tilde{q}_{\text{2D}}) \left\{ r/w_1-3/ r^2 \right\}$. The ratio of the near-resonant inelastic rate constant to the off-resonant one is ${\cal R}_{\text{2D}} \sim (\psi_{\text{2D}}^{\text{res}} / \psi_{\text{2D}}^{\text{off}})^2 F_{\text{2D}}$, where we have to put $r \sim R_e$ in the expressions for the relative wavefunctions and take into account that in the off-resonant regime $w_1\sim R_e^3$, whereas in the near-resonant regime it is much larger. This yields \cite{noteRes}
\be \label{R_2D}
	{\cal R}_{\text{2D}}\equiv \left< \alpha_{\text{2D}}^{\text{res}} \right>_T / \left< \alpha_{\text{2D}}^{\text{off}} \right>_T \sim \frac{l_0}{R_e}\frac{1}{(q_T R_e)^4},
\ee
which is consistent with equations (\ref{alpha_2D_highT_off_res}) and (\ref{alpha_2D_highT_res}). As one can see from Eqs. (\ref{R_3D}) and (\ref{R_2D}), the ratio ${\cal R}_{\text{2D}}/{\cal R}_{\text{3D}} \sim l_0 k_T\ll 1$. Thus, in 2D the enhancement of the inelastic rate constant near the resonance is suppressed compared to 3D.

\begin{widetext}

\begin{figure}
\subfloat{%
  \includegraphics[width=.49\linewidth]{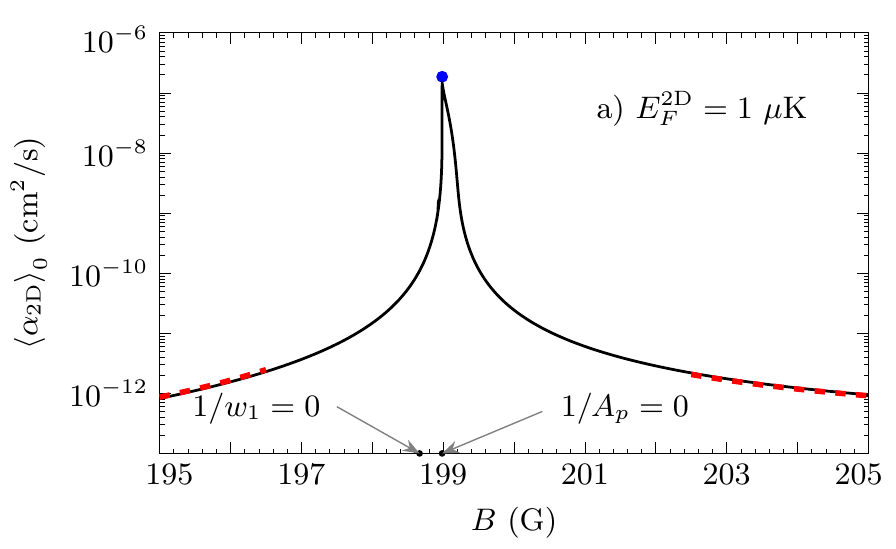}%
}\hfill
\subfloat{%
  \includegraphics[width=.49\linewidth]{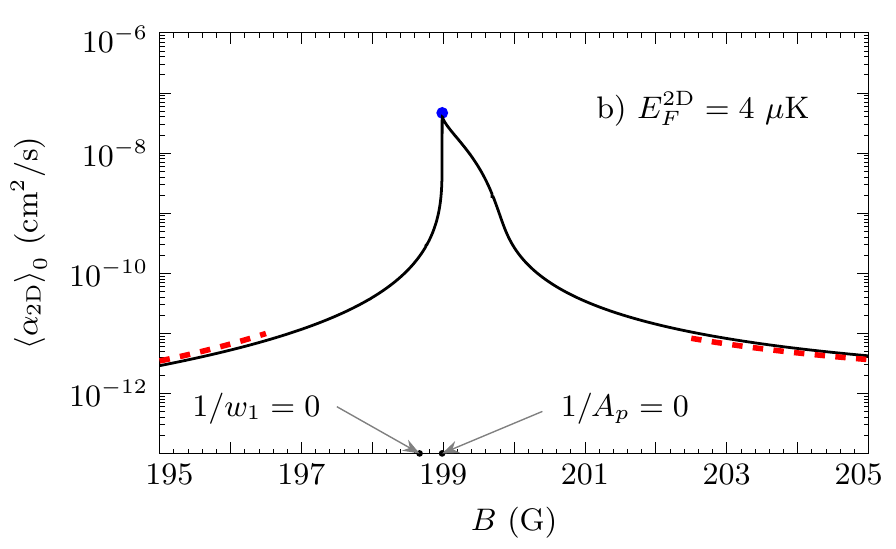}%
}
\caption{Two-dimensional inelastic rate constant $\left<\alpha_{\text{2D}}\right>_0$ for $^{40}$K atoms in the $\ket{9/2,-7/2}$ state at $T = 0$ versus magnetic field~$B$ for $E^{\text{2D}}_F = 1$ $\mu$K in a) and $E^{\text{2D}}_F = 4$ $\mu$K in b). Dashed red curves correspond to the off-resonant regime described by Eq.~(\ref{alpha_2D_zeroT_off_res}), and the blue point marks the near-resonant peak value according to Eq. (\ref{alpha_2D_zeroT_res}). It is shifted in the direction of higher fields by $1$~mG in a) and by $1.9$ mG in b) with respect to the magnetic field at which $1/A_p = 0$. The confining frequency is $\omega_0 = 120$~kHz.}
\label{alpha2D_ZeroT}
\end{figure}

\begin{figure}
\subfloat{%
  \includegraphics[width=.49\linewidth]{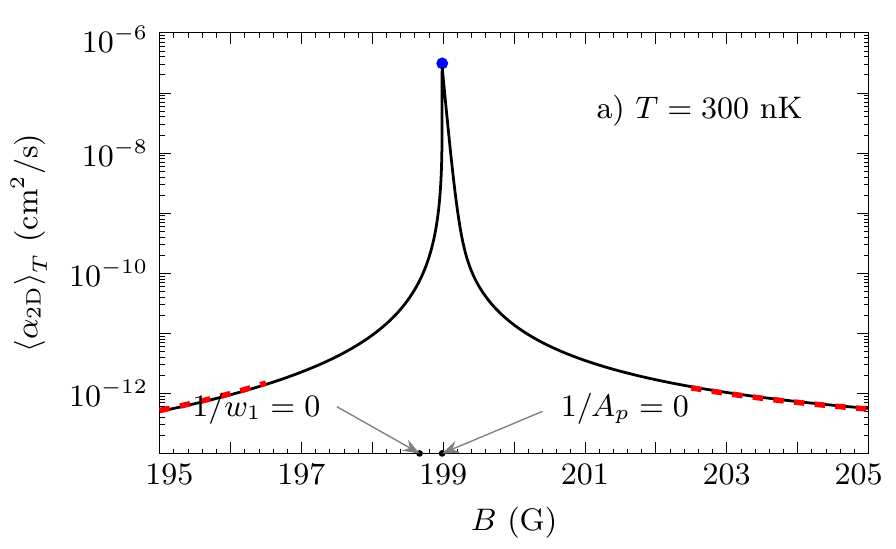}%
}\hfill
\subfloat{%
  \includegraphics[width=.49\linewidth]{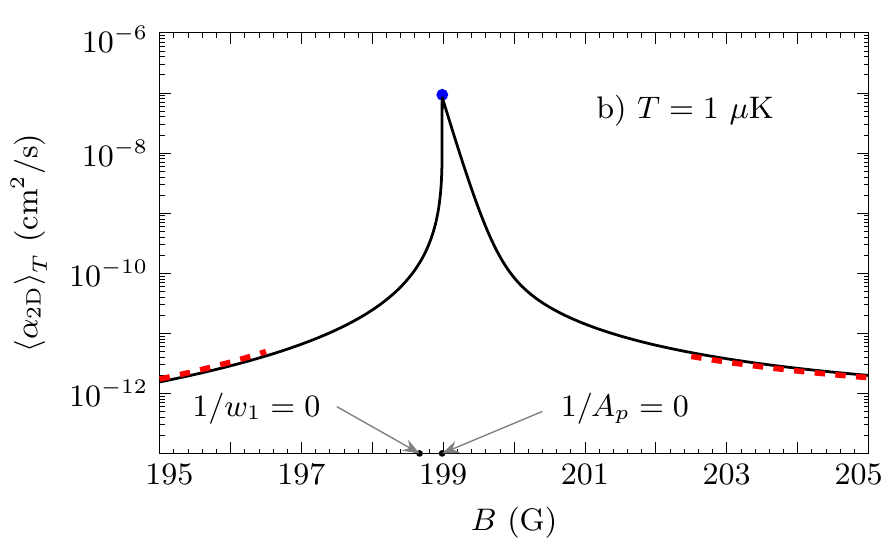}%
}
\caption{Two-dimensional inelastic rate constant $\left<\alpha_{\text{2D}}\right>_T$ for $^{40}$K atoms in the $\ket{9/2,-7/2}$ state versus magnetic field $B$ for $T = 300$ nK in a) and $T = 1$ $\mu$K in b). Dashed red curves correspond to the off-resonant regime described by Eq. (\ref{alpha_2D_highT_off_res}), and the blue point marks the near-resonant peak value according to Eq. (\ref{alpha_2D_highT_res}). It is shifted in the direction of higher fields by $1.5$~mG in a) and by $2.7$ mG in b) with respect to the magnetic field at which $1/A_p = 0$. The confining frequency is $\omega_0 = 120$ kHz.}
\label{alpha2D_HighT}
\end{figure}

\end{widetext}

\section{2-body inelastic collisions in 1D} \label{section1D}

We eventually turn to inelastic collisions in the one-dimensional case. Omitting the doubling of the resonance, induced by the spin-dipole interaction, we have a single-peak structure of the relaxation rate constant. This structure is realized for the magnetic field parallel to the line of the translational motion or for the field perpendicular to this line \cite{Gunter2005, Peng2014}. The shift of the peak due to the spin-dipole interaction is discussed in the Appendix. We also present there the calculations taking into account the spin-dipole doubling of the resonance and the resulting double-peak structure of the relaxation rate for the magnetic field forming the angle of $45\degree$ with the line of the translational motion.
  
In the quasi-1D geometry obtained by a tight harmonic confinement in two directions ($x,y$) with frequency $\omega_0$, the wavefunction of the relative motion in the odd-wave channel (analog of $p$-wave in 2D and 3D) is
\be
	\psi_{\text{1D}}({\bf r}) = \chi_{\text{1D}}(z) \frac{1}{\sqrt{2\pi}l_{0}} \exp{\left\{ -\frac{\rho^2}{4l_{0}^2} \right\}},
\ee
where $z$ is the longitudinal interparticle separation, $\rho =\sqrt{x^2+y^2}$ is the transverse separation, and  $l_{0} = \sqrt{\hbar/(m \omega_{0})}$ is the transverse extension of the wavefunction.
The longitudinal motion with the 1D relative momentum $q$ at distances $|z| \gg l_0 \gg R_e$ is described by the wavefunction
\be \label{chi_1D}
	\chi_{\text{1D}}(z) = i \sin qz + \text{sgn}(z) f_{\text{1D}}(q) e^{i q |z|},
\ee
with the odd-wave scattering amplitude $f_{\text{1D}}(q)$ given by \cite{Pricoupenko2008, Peng2014, Gao2015}
\be \label{scatt_amp_1D}
	f_{\text{1D}}(q) = \frac{-i q}{1/l_p + \xi_p q^2 + i q},
\ee
where $l_p = 3l_{0} \left[ l_{0}^3/w_1 + \alpha_1 l_0 + 3\sqrt{2}|\zeta(-1/2)| \right]^{-1}$ and $\xi_p = \alpha_1 l_{0}^2/3$ are the 1D scattering parameters, and $\zeta(-1/2) \approx -0.208$ is the Riemann zeta-function. The amplitude $f_{\text{1D}}(k)$ is related to the 1D odd-wave S-matrix element $S_{\text{1D}}(q)$ as  $f_{\text{1D}}(q) = (S_{\text{1D}}(q) - 1)/2$. Like in higher dimensions, in the presence of inelastic processes we should replace $1/w_1$ with $1/w_1 + i/w'_1$, which gives the following expression for the S-matrix element:
 \be
 	S_{\text{1D}}(q) = \frac{1/l_p + \xi_p q^2 + i (1/l'_p - q)}{1/l_p + \xi_p q^2 + i (1/l'_p + q)},
 \ee
 where $1/l'_p = l_{0}^2 / 3 w'_1 >0$. Then, writing the wavefunction (\ref{chi_1D}) at $|z| \to \infty$ as $\chi_{\text{1D}} = \text{sgn}(z) (1/2) \left\{ -\exp{(-i q |z|)} + S_{\text{1D}}(q) \exp{(i q |z|)} \right\}$, we see that the intensity of the outgoing wave is reduced by a factor of $|S_{\text{1D}}(q)|^2 < 1$ compared to the incoming wave. The inelastic cross-section in 1D is defined as $\sigma_{\text{1D}}^{\text{in}} = \left(1-|S_{\text{1D}}(q)|^2 \right)/2$, and for identical particles there is an additional factor of 2. Then, for the inelastic rate constant, $\alpha_{\text{1D}}(q) = (2\hbar q / m) \sigma_{\text{1D}}^{\text{in}}$, we obtain:
 \be \label{alpha_1D}
 	\alpha_{\text{1D}}(q) = \frac{8\hbar}{m l'_p}\frac{q^2}{\left[ 1/l_p + \xi_p q^2 \right]^2 + \left[ 1/l'_p + q \right]^2}.
\ee
Sufficiently far from the 1D resonance ($1/l_p=0$), where the dominant term in the denominator of Eq. (\ref{alpha_1D}) is $1/l_p$, the rate constant becomes
 \be \label{alpha_1D_off_res}
 	\alpha_{\text{1D}}(q) \approx \frac{8\hbar}{m}\frac{l_p^2}{l'_p} q^2 \approx \frac{24 \hbar}{m l_{0}^2} \frac{w_1^2}{w'_1} q^2.
 \ee
 
At $T = 0$ the off-resonant rate constant averaged over the Fermi step momentum distribution reads
\be \label{alpha_1D_zeroT_off_res}
	\left< \alpha_{\text{1D}} \right>_0 \approx \frac{8 w_1^2}{l_{0}^2 w'_1}\frac{E^{\text{1D}}_F}{\hbar},
\ee
where $E^{\text{1D}}_F = \hbar^2 q_F^2/2m$ is the 1D Fermi energy, $q_F = \pi n_{\text{1D}}$ is the Fermi momentum for a single-component 1D gas and $n_{\text{1D}}$ is the 1D density. 

In the vicinity of the 1D resonance on its negative side $(l_p < 0)$ the largest contribution to the rate constant comes from momenta $\sim \tilde{q}_{\text{1D}} = 1/\sqrt{\xi_p |l_p|}$. Then, in the near-resonant regime, where $1/l'_p \ll \tilde{q}_{\text{1D}} \ll q_F$, the rate constant shows a narrow peak with the value
\be \label{alpha_1D_zeroT_res}
	\left< \alpha_{\text{1D}} \right>_0 \approx \frac{8 \pi \hbar}{l'_p \xi_p \, m}\frac{1}{q_F} = \frac{8 \pi \hbar}{\alpha_1 m \, w'_1}\frac{1}{q_F}.
\ee
As in the 2D case, the maximum value of $\left< \alpha_{\text{1D}} \right>_0$ is almost independent of the confinement frequency.

The results of direct numerical calculation of $\left< \alpha_{\text{1D}} \right>_0$ using Eq. (\ref{alpha_1D}) for $^{40}$K atoms are presented in Fig. \ref{alpha1D_ZeroT} for the confining frequency $\omega_0 = 120$ kHz and Fermi energies $E^{\text{1D}}_F = 1$ $\mu$K and $4$ $\mu$K (corresponds to densities $n_{\text{1D}} \approx 4.1 \times 10^{4}$ cm$^{-1}$ and $8.2 \times 10^{4}$ cm$^{-1}$, respectively). The off-resonant expression (\ref{alpha_1D_zeroT_off_res}) and near-resonant expression~(\ref{alpha_1D_zeroT_res}) agree with numerical results, although Eq.~(\ref{alpha_1D_zeroT_res}) leads to a small overestimate of $\left< \alpha_{\text{1D}} \right>_0$.

At $T \gg E^{\text{1D}}_F$, averaging the rate constant over the Boltzmann distribution of atoms we obtain the following off-resonant expression:
\be \label{alpha_1D_highT_off_res}
	\left< \alpha_{\text{1D}} \right>_T \approx \frac{12 w_1^2}{l_{0}^2 w'_1}\frac{T}{\hbar}.
\ee

On the negative side of the 1D resonance in the near-resonant regime, where $1/l'_p \ll \tilde{q}_{\text{1D}} \ll q_T$, with $q_T = \sqrt{m T /\hbar^2}$ being the thermal momentum, the rate constant displays a sharp peak, slightly shifted with respect to the position of the 1D resonance at zero kinetic energy ($1/l_p = 0$). The maximum value of the rate constant is
\be \label{alpha_1D_highT_res}
\begin{aligned}
	\left< \alpha_{\text{1D}} \right>_T \approx \frac{4 \sqrt{\pi} \hbar}{l'_p \xi_p \, m}\frac{1}{q_T} = \frac{4 \sqrt{\pi} \hbar}{\alpha_1 m \, w'_1}\frac{1}{q_T}.
\end{aligned}
\ee

Direct numerical calculation of $\left< \alpha_{\text{1D}} \right>_T$ on the basis of Eq.~(\ref{alpha_1D}) shows good agreement with both off-resonant and near-resonant expressions, as shown in Fig.~\ref{alpha1D_HighT} for the confining frequency $\omega_{0} = 120$ kHz and temperatures $T = 300$ nK and $T = 1$ $\mu$K. Note that in the vicinity of the peak value the rate constant is proportional to $1/\sqrt{T}$, while in the off-resonant regime it has a linear dependence on $T$.

We see that in 1D, as well as in higher dimensions, the temperature (Fermi energy) dependence of the inelastic rate constant in the near-resonant regime is very different from that in the off-resonant case. Similarly to the 2D case, at distances where the relaxation occurs ($R_e \lesssim r \ll l_0$), the 1D relative wavefunction $\psi_{\text{1D}}$ has a 3D character and differs from the 3D wavefunction only by a normalization coefficient. Sufficiently far from the 3D resonance, assuming that $r$ is still larger than $R_e$, the 1D wavefunction can be written as \cite{Pricoupenko2008}
\be \label{psi_1D_small_distances}
	\psi_{\text{1D}}(r) = -\frac{f_{\text{1D}}(q)\sqrt{2\pi}l_0}{6\pi w_1} \left\{ r - 3w_1/r^2 \right\}.
\ee
Far from the 1D resonance ($1/l_p = 0$) the 1D scattering amplitude is $f_{\text{1D}}^{\text{off}} \approx - 3 i w_1 q / l_0^2$, which yields $\psi_{\text{1D}}^{\text{off}} \approx (i q /\sqrt{2\pi}l_0) \left\{ r - 3w_1 / r^2 \right\}$ and $\alpha_{\text{1D}}^{\text{off}} \sim q^2 / l_0^2$, in agreement with Eq. (\ref{alpha_1D_off_res}). In the near-resonant regime on the negative side of the confinement-influenced resonance ($l_p < 0$) the situation changes. Here the main contribution to $\alpha_{\text{1D}}$ is provided only by relative momenta in a narrow interval $\delta q \sim 1 / \xi_p$ around $\tilde{q}_{\text{1D}}$, and in the classical gas the fraction of such momenta is $F_{\text{1D}} \sim \delta q / q_T \sim 1 /(\xi_p q_T)$. In this near-resonant regime we have $|1/l_p + \xi_p q^2| \sim \tilde{q}_{\text{1D}}$ and $f_{\text{1D}}^{\text{res}}(q) \approx -1$. Then, the 1D wavefunction becomes $\psi_{\text{1D}}^{\text{res}} \sim l_0\left\{ r/w_1-3/ r^2 \right\}$. The ratio of the near-resonant inelastic rate constant to the off-resonant one is ${\cal R}_{\text{1D}} \sim (\psi_{\text{1D}}^{\text{res}} / \psi_{\text{1D}}^{\text{off}})^2 F_{\text{1D}}$, where we have to put $r \sim R_e$ in the expressions for the relative wavefunctions. Taking into account that in the off-resonant regime $w_1\sim R_e^3$ and in the near-resonant regime it is much larger, we obtain:
\be \label{R_1D}
	{\cal R}_{\text{1D}}\sim \left( \frac{l_0}{R_e} \right)^2 \frac{1}{(q_T R_e)^3},
\ee
which is consistent with equations (\ref{alpha_1D_highT_off_res}) and (\ref{alpha_1D_highT_res}). From Eqs. (\ref{R_3D}), (\ref{R_2D}), and (\ref{R_1D}) we find that ${\cal R}_{\text{1D}}/{\cal R}_{\text{3D}} \sim (k_T l_0)^2 \sim (k_T l_0) {\cal R}_{\text{2D}}/ {\cal R}_{\text{3D}}$. Thus, in the 1D case the enhancement of the inelastic rate near the resonance is even weaker than in 2D and certainly much weaker than in 3D.

\begin{widetext}

\begin{figure}
\subfloat{%
  \includegraphics[width=.49\linewidth]{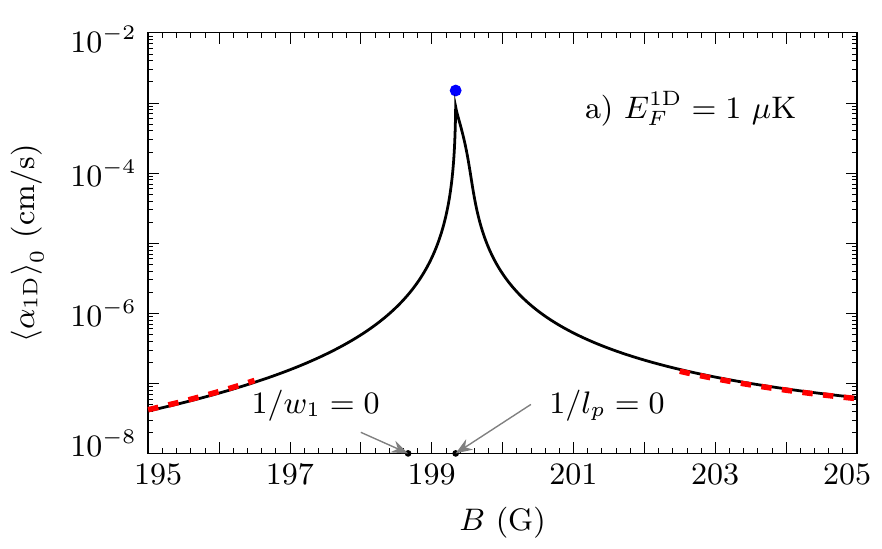}%
}\hfill
\subfloat{%
  \includegraphics[width=.49\linewidth]{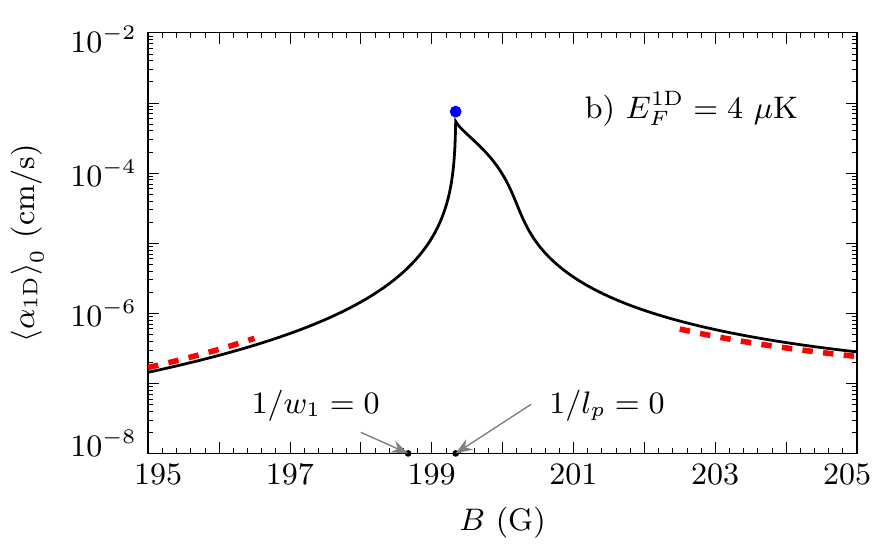}%
}
\caption{One-dimensional inelastic rate constant $\left<\alpha_{\text{1D}}\right>_0$ for $^{40}$K atoms in the $\ket{9/2,-7/2}$ state at $T = 0$ versus magnetic field $B$ for $E^{\text{1D}}_F = 1$ $\mu$K in a) and $E^{\text{1D}}_F = 4$ $\mu$K in b). Dashed red curves correspond to the off-resonant regime described by Eq. (\ref{alpha_1D_zeroT_off_res}), and the blue point marks the near-resonant peak value according to Eq. (\ref{alpha_1D_zeroT_res}). The corresponding magnetic field practically coincides with the magnetic field at which $1/l_p = 0$. The confining frequency is $\omega_0 = 120$ kHz.}
\label{alpha1D_ZeroT}
\end{figure}

\begin{figure}
\subfloat{%
  \includegraphics[width=.49\linewidth]{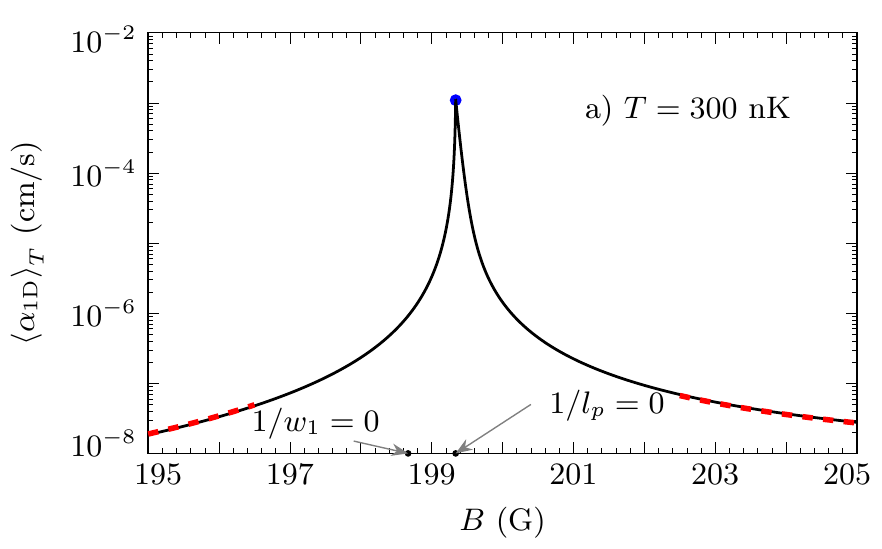}%
}\hfill
\subfloat{%
  \includegraphics[width=.49\linewidth]{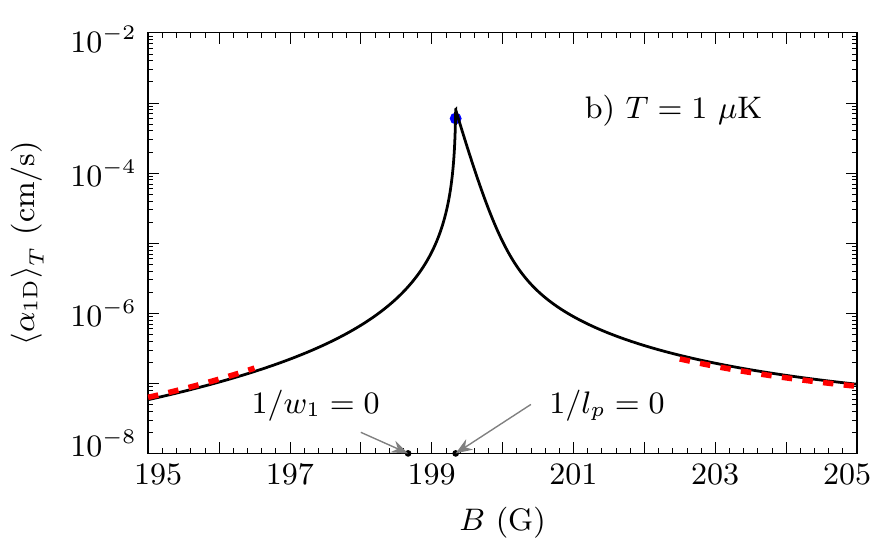}%
}
\caption{One-dimensional inelastic rate constant $\left<\alpha_{\text{1D}}\right>_T$ for $^{40}$K atoms in the $\ket{9/2,-7/2}$ state versus magnetic field $B$ for $T = 300$ nK in a) and $T = 1$ $\mu$K in b). Dashed red curves correspond to the off-resonant regime described by Eq. (\ref{alpha_1D_highT_off_res}), and the blue point marks the near-resonant peak value according to Eq. (\ref{alpha_1D_highT_res}). The corresponding magnetic field practically coincides with the magnetic field at which $1/l_p = 0$. The confining frequency is $\omega_0 = 120$ kHz.}
\label{alpha1D_HighT}
\end{figure}

\end{widetext}


\section{2-body inelastic rate near the resonance in 2D and 1D. Conclusions}

In this section we analyze how the inelastic rate is enhanced on approach to the resonance in reduced dimensionalities and conclude. 

In order to demonstrate the suppressed enhancement of the inelastic rate constant near the resonance in reduced dimensionalities, we calculate the ratios of $\left< \alpha_{\text{2D}} \right>_T$ and $\left< \alpha_{\text{1D}} \right>_T$ to their off-resonant values, and compare them with the ratio of $\left< \alpha_{\text{3D}} \right>_T$ to its value far from the resonance. In Fig. \ref{2Dvs3D_HighT} we plot the ratio of the 2D rate constant to its off-resonant value, $\left< \alpha_{\text{2D}} \right>_T / \left< \alpha_{\text{2D}}^{\text{off}} \right>_T$, versus magnetic field $B$ for $^{40}$K atoms in the $\ket{9/2,-7/2}$ state at $T=300$ nK and $T=1$ $\mu$K. The off-resonant value is taken at a fixed field value of $195$ G. Fig. \ref{1Dvs3D_HighT} shows the corresponding quantity in 1D. It is evident that the rate constant in 3D experiences a much stronger enhancement near the resonance than the rate constants in 2D and 1D. In other words, this means that the enhancement of the 2-body inelastic rate near the resonance is suppressed in reduced dimensionalities. The effect is especially pronounced in 1D, which is consistent with our discussion in the previous section. 

This effect is mostly related to a weaker enhancement of the relative wavefunction on approach to the resonance in 2D and 1D than in 3D. Indeed, using expressions for $\psi_{\text{3D}}$ and $\psi_{\text{2D}}$ in the near- and off-resonant regimes (written after Eqs. (\ref{psi_3D_small_distances}) and (\ref{psi_2D_small_distances})) we see that the ratio $\left( \psi_{\text{2D}}^{\text{res}} / \psi_{\text{2D}}^{\text{off}} \right)^2 / \left( \psi_{\text{3D}}^{\text{res}} / \psi_{\text{3D}}^{\text{off}} \right)^2 \sim (\tilde k_{\text{3D}} l_0)^2 \ll 1$ slightly away from the 3D resonance. Similarly, in 1D we have $\left( \psi_{\text{1D}}^{\text{res}} / \psi_{\text{1D}}^{\text{off}} \right)^2 / \left( \psi_{\text{3D}}^{\text{res}} / \psi_{\text{3D}}^{\text{off}} \right)^2 \sim (\tilde k_{\text{3D}} l_0)^4$, which is even smaller than in the 2D case.

Our results may draw promising paths to obtain novel many-body states in 2D and 1D, such as low density $p$-wave (odd-wave) superfluids of spinless fermions. It is quite likely that they can be extended to the case of three-body recombination \cite{note3}, which will be the topic of our future research.

\begin{widetext}

\begin{figure}[H]
\subfloat{%
  \includegraphics[width=.49\linewidth]{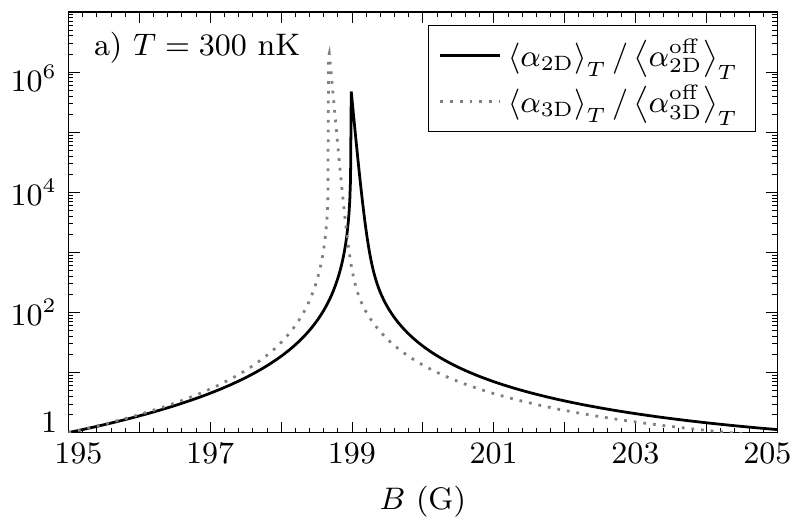}%
}\hfill
\subfloat{%
  \includegraphics[width=.49\linewidth]{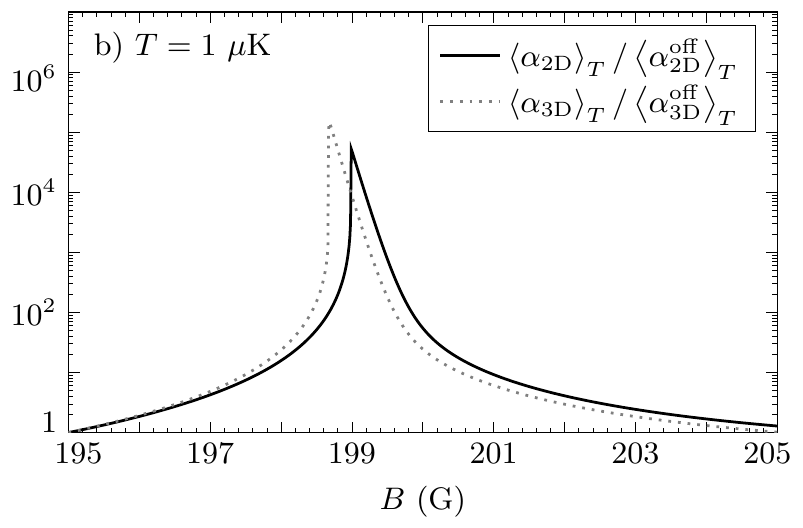}%
}
\caption{Inelastic rate constants in 2D (solid curve) and in 3D (dotted curve) divided by their off-resonant values at a fixed filed of 195 G for $^{40}$K atoms in the $\ket{9/2,-7/2}$ state versus magnetic field $B$ for $T = 300$ nK in a) and $T = 1$ $\mu$K in b).}
\label{2Dvs3D_HighT}
\end{figure}

\begin{figure}[H]
\subfloat{%
  \includegraphics[width=.49\linewidth]{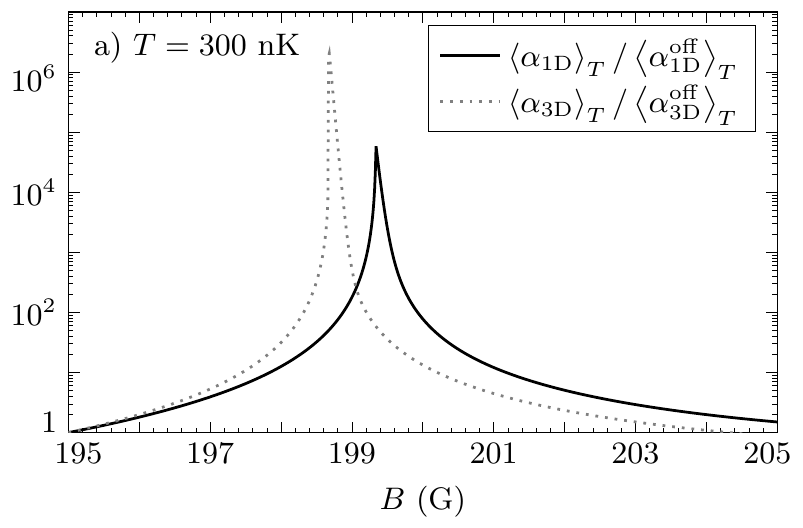}%
}\hfill
\subfloat{%
  \includegraphics[width=.49\linewidth]{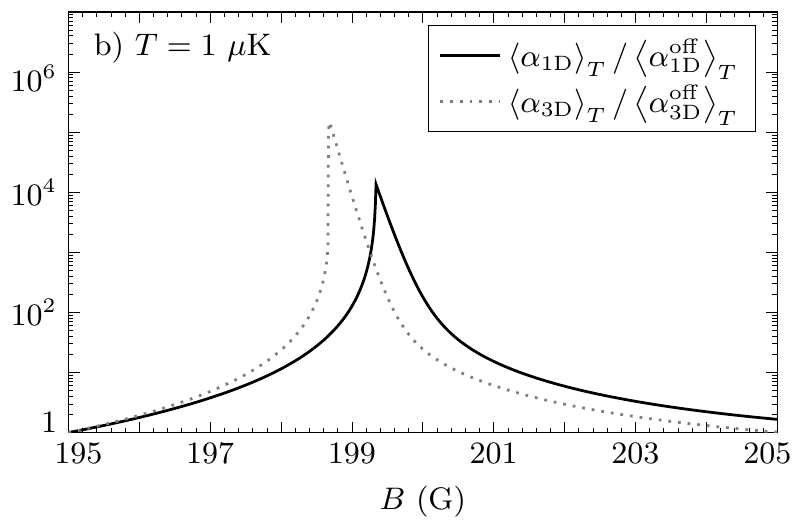}%
}
\caption{Inelastic rate constants in 1D (solid curve) and in 3D (dotted curve) divided by their off-resonant values at a fixed filed of 195 G for $^{40}$K atoms in the $\ket{9/2,-7/2}$ state versus magnetic field $B$ for $T = 300$ nK in a) and $T = 1$ $\mu$K in b).}
\label{1Dvs3D_HighT}
\end{figure}

\end{widetext}

\begin{acknowledgments}
We would like to thank D. Petrov and A. Fedorov for useful discussions and J. Bohn for providing us with the results of coupled-channel numerical calculations. This research was supported in part by the National Science Foundation under Grant No. NSF PHY-1125915. GVS is grateful to the Kavli Institute
of Theoretical Physics for hospitality during the workshop on Universality of Few-Body Systems (November-December, 2016), where part of the work has been done. We also acknowledge support from IFRAF and from the Dutch Foundation FOM. The research leading to these results has received funding from the European Research Council under European Community's Seventh Framework Programme (FR7/2007-2013 Grant Agreement no. 341197).
\end{acknowledgments}

\indent
\appendix*
\section{Spin-relaxation taking into account the doubling of the $p$-wave Feshbach resonance}
Throughout the paper we assumed that the 3D $p$-wave Feshbach resonance occurs at the same magnetic field for all orbital angular momentum projections $m_l$. However, in reality due to the spin-dipole interaction between colliding atoms the binding energy of the two-body bound state, the coupling to which leads to the resonance in the scattering amplitude, depends on $|m_l|$. As a consequence, the resonant magnetic field (at which the scattering volume diverges) is different for $m_l = 0$ and $m_l = \pm1$. Then the 3D rate constant exhibits a doublet structure: there are two distinct peaks, corresponding to $m_l = 0$ and $|m_l| = 1$. In reduced dimensionalities one can also get a double-peak structure of the relaxation rate constant, although the situation is more peculiar as the orientation of the external magnetic field plays a crucial role\cite{Gunter2005, Peng2014, Gao2015}. In the Appendix we analyze these effects in more detail. We first derive an expression for the $m_l$-dependent inelastic rate constant in 3D and show that it has the expected doublet structure. We then discuss how our results for the rate constants in 2D and 1D are affected by the $m_l$-dependence of the $p$-wave Feshbach resonance.

\begin{figure}[H]
	\includegraphics[width=\linewidth]{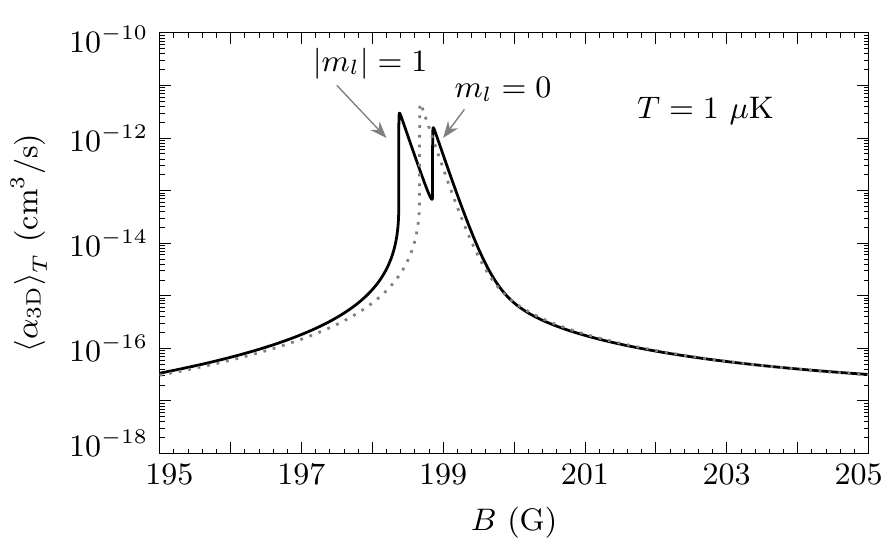}%
\caption{Three-dimensional inelastic rate constant $\left<\alpha_{\text{3D}}\right>_T$ for $^{40}$K atoms in the $\ket{9/2,-7/2}$ state versus magnetic field $B$ for $T = 1$ $\mu$K. For comparison, the grey dotted line shows the 3D inelastic rate constant from Fig. \ref{alpha3D_HighT} b) of the main text.}
\label{alpha3D_HighT_doublet}
\end{figure}

In 3D, if the scattering volume and the effective range depend on the value of $m_l$, then the $p$-wave scattering phase shift also becomes $m_l$-dependent. In the low energy limit we have $k^3 \cot \delta_{m_l} = -1/w_{1,m_l} - \alpha_{1,m_l}$, and the $p$-wave part of the total scattering amplitude can be written as \cite{LL}
\be \label{scatt_amp_aniso}
	f(\boldsymbol{k}, \hat{\boldsymbol{k}}') = 4\pi \sum_{m_l = 0,\pm 1} f_{m_l}(k) Y_{1,m_l}^*(\hat{\boldsymbol{k}})Y_{1,m_l}(\hat{\boldsymbol{k}}'),
\ee
where $\hat{\boldsymbol{k}}$ and $\hat{\boldsymbol{k}}'$ are unit vectors in the directions of incident and outgoing relative momenta, $Y_{1,m_l}$ is the spherical harmonic, and $f_{m_l}(k) = (S_{m_l}(k) - 1)/2ik$ is the $p$-wave partial scattering amplitude, with $S_{m_l}(k) = \exp\left\{ 2i\delta_{m_l} \right\}$ being the $p$-wave S-matrix element. In order to describe inelastic collisions we make the replacement $1/w_{1,m_l} \to 1/w_{1,m_l} + 1/w'_1$, where $w'_1 > 0$ is the same for all $m_l$'s, since it can be assumed to be field-independent. From Eq. (\ref{scatt_amp_aniso}) we see that the $p$-wave part of the scattering amplitude depends on both the incoming and outgoing momentum directions and not only on the angle between $\boldsymbol{k}$ and $\boldsymbol{k}'$ (as it would have been in the case where the scattering phase shift is independent of $m_l$). Then, integrating over the scattering angles $\Omega_{\hat{\boldsymbol{k}}'}$, for the inelastic scattering cross section we have:
\be \label{cross_section_aniso}
	\sigma^{\text{in}}(\boldsymbol{k}) = \frac{4 \pi}{k^2} \sum_{m_l = 0, \pm 1}  \left[ 1 - \left| S_{m_l}(k) \right|^2 \right] \left| Y_{1,m_l}(\hat{\boldsymbol{k}}) \right|^2.
\ee 
For identical particles the above expression should be multiplied by an extra factor of 2. Taking into account that $|Y_{1,0}(\hat{\boldsymbol{k}})|^2 = (3/4\pi) \cos \theta_{\hat{\boldsymbol{k}}}$ and $|Y_{1,\pm1}(\hat{\boldsymbol{k}})|^2 = (3/8\pi) \sin^2 \theta_{\hat{\boldsymbol{k}}}$, where $\theta_{\hat{\boldsymbol{k}}}$ is the angle between the unit vector $\hat{\boldsymbol{k}}$ and the quantization axis, we average expression (\ref{cross_section_aniso}) over the incident angles $\Omega_{\hat{\boldsymbol{k}}}$. Then, the inelastic cross section can be written as $\bar \sigma^{\text{in}}(k) = \sum_{m_l = 0, \pm 1} \bar \sigma^{\text{in}}_{m_l}(k)$, where $\bar \sigma^{\text{in}}_{m_l}(k) = (2\pi/k^2)[ 1 - | S_{m_l}(k) |^2 ]$.
Accordingly, for the inelastic rate constant in 3D, $\alpha_{\text{3D}}(k) = (2\hbar k/m) \bar \sigma^{\text{in}}(k)$, we obtain:
\be \label{alpha_3D_appendix}
\begin{aligned}
	&\alpha_{\text{3D}}(k) = \frac{16\pi\hbar}{m w'_1} \, \times\\
	& \mathlarger{\sum_{m_l = 0,\pm1}} \frac{k^2}{\left[ \mathlarger{\frac{1}{w_{1,m_l}}} + \alpha_{1,m_l} k^2 \right]^2 + \left[ \mathlarger{\frac{1}{w'_1}} + k^3 \right]^2}.
\end{aligned}
\ee
One immediately sees that if $w_{1,m_l}$ and $\alpha_{1,m_l}$ are the same for all $m_l$'s, the above expression reduces to Eq. (\ref{alpha_3D}) of the main text, which has only one peak. However, as we already mentioned before, if one takes the spin-dipole doubling of the resonance into account, then this peak splits in two smaller peaks. The one which corresponds to $|m_l|=1$ is by a factor of $3/2$ smaller, whereas the second peak, corresponding to $m_l = 0$, is smaller by a factor of $3$. For $^{40}$K atoms in the $\left| 9/2, -7/2 \right>$ state the 3D resonance for $m_l = 0$ occurs at 198.8 G, and for $|m_l| = 1$ at 198.3 G \cite{Ticknor2004}. We present this in Fig. \ref{alpha3D_HighT_doublet}, which displays the thermally averaged inelastic rate constant $\left< \alpha_{\text{3D}} \right>_T$ in a 3D classical gas at 1 $\mu$K.

\begin{widetext}

\begin{figure}[H]
\subfloat{%
  \includegraphics[width=.49\linewidth]{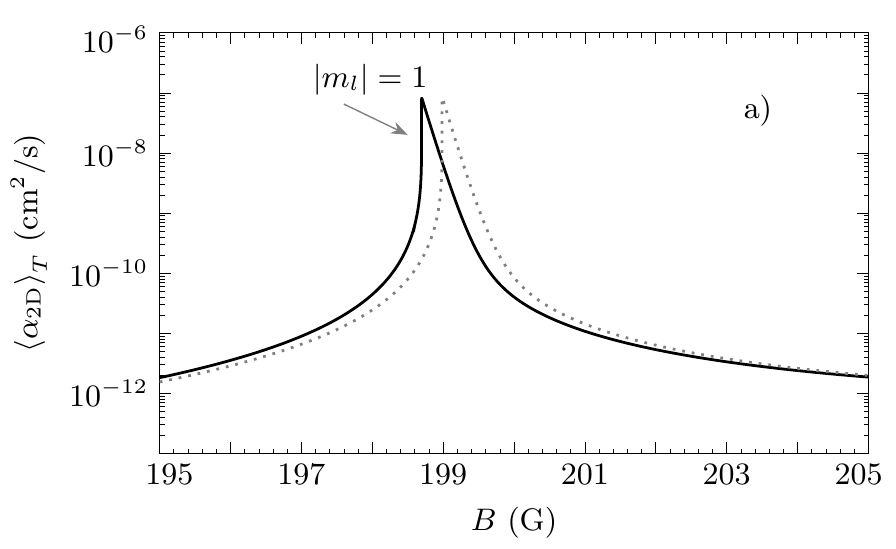}%
}\hfill
\subfloat{%
  \includegraphics[width=.49\linewidth]{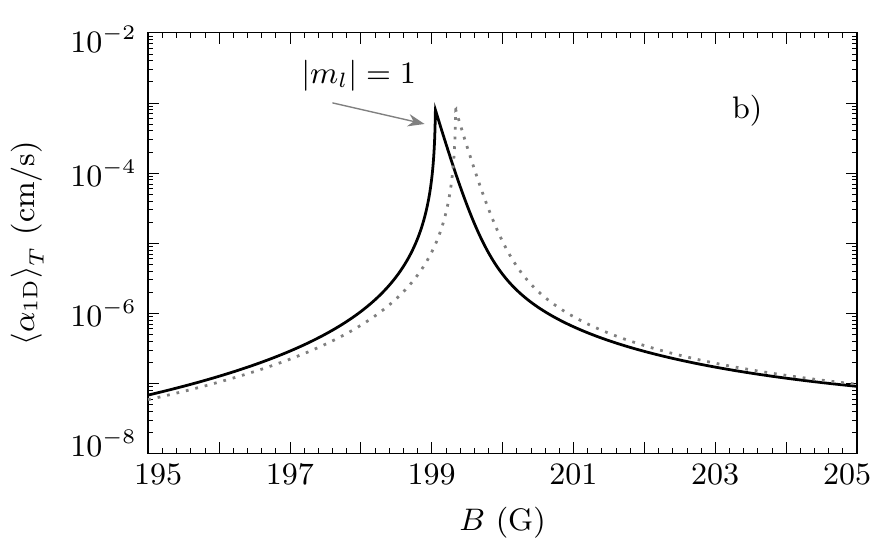}%
}
\caption{Two-dimensional inelastic rate constant $\left<\alpha_{\text{2D}}\right>_T$ in a) and one-dimensional inelastic rate constant $\left<\alpha_{\text{1D}}\right>_T$ in b) for $^{40}$K atoms in the $\ket{9/2,-7/2}$ state versus magnetic field $B$ for $T = 1$ $\mu$K and confining frequency $\omega_0 = 120$ kHz. The magnetic field is perpendicular to the translational motion in both a) and b). For comparison, grey dotted lines show the corresponding inelastic rate constants from Fig. \ref{alpha2D_HighT} b) and Fig. \ref{alpha1D_HighT} b) of the main text. One can see that the spin-dipole interaction shifts the peaks by approximately 0.3 G.}
\label{alpha2D1D_HighT_ml1}
\end{figure}

\end{widetext}

In reduced dimensionalities the two-body inelastic relaxation occurs at interparticle distances that are much smaller than the extension of the relative wavefunction in the tightly confined direction(s) \cite{Petrov2001}. Therefore, the relative motion acquires a 3D character and the related wavefunction represents a superposition of $m_l=0$ and $m_l=\pm 1$ contributions. This means that in principle the rate constant in 2D and in 1D can also have the double-peak structure. However, the number of peaks and their positions depend on the relative orientation of the external magnetic field \cite{Gunter2005, Peng2014}. In the following, we first consider the quasi-2D case with the magnetic field perpendicular to the plane of the translational motion. In quasi-1D we assume that the field is perpendicular to the line of the translational motion. In both cases one has a single-peak structure of the relaxation rate, since the relative wavefunction of two atoms at short separations corresponds only to the 3D motion with $|m_l|=1$. The expressions for quasi-2D and quasi-1D scattering amplitudes for an arbitrary orientation of the magnetic field were derived in Ref. \cite{Peng2014}. In the case of magnetic field perpendicular to the plane (line) of the translational motion in 2D (1D), these expressions are reduced to Eqs. (\ref{scatt_amp_2D}) and (\ref{scatt_amp_1D}) for $f_{\text{2D}}$ and $f_{\text{1D}}$, correspondingly, where one should use the values of $w_1$ and $\alpha_1$ for $|m_l| = 1$. Thus, the spin-dipole interaction simply shifts the position of the peak of the inelastic rate by approximately $0.3$ G, as can be seen from Fig. \ref{alpha2D1D_HighT_ml1}. For the 1D case with the magnetic field perpendicular to the line of the translational motion, particle losses in a spin-polarized gas of $^{40}$K atoms were measured in Ref. \cite{Gunter2005}. The peak position observed in the experiment coincides with that in Fig. \ref{alpha2D1D_HighT_ml1}~b).

To illustrate the effect of the spin-dipole doubling of the resonance in reduced dimensionalities, we now turn to the case where the magnetic field is parallel to the plane of the translational motion in 2D. Using the result  of Ref. \cite{Peng2014} for the 2D scattering amplitude depending on the orientation of the $B$-field, the inelastic scattering cross section can be written as
\be \label{sigma_2D_appendix}
\begin{aligned}
	\sigma_{\text{2D}}^{\text{in}}(\boldsymbol{q}) = \frac{4}{q}\Bigl\{ &\left[ 1- \left| S_0^{\text{2D}} \right|^2 \right] \cos^2 \phi_{\hat{\boldsymbol{q}}} \Bigr. \\
	& \Bigl. + \left[ 1- \left| S_1^{\text{2D}} \right|^2 \right] \sin^2 \phi_{\hat{\boldsymbol{q}}} \Bigr\},
\end{aligned}
\ee
where $\boldsymbol{q}$ is the incident relative momentum, $\phi_{\hat{\boldsymbol{q}}}$ is the angle between $\boldsymbol{q}$ and the external magnetic field, and $S_{|m_l|}^{\text{2D}} = \exp \left\{ 2 i \delta_{|m_l|}^{\text{2D}}\right\}$ are the 2D $p$-wave S-matrix elements, with $\delta_{|m_l|}^{\text{2D}}$ being the 2D $p$-wave $|m_l|$-dependent scattering phase shifts. Adopting our notations from the main text, in the low energy limit we can write $q^2 \cot \delta_{|m_l|}^{\text{2D}} = -1/A_{p, |m_l|} + \left[B_{p,|m_l|} -(2/\pi)\ln l_0 q\right] q^2$. The quantities $A_{p,|m_l|}$ and $B_{p, |m_l|}$ are given by the same expressions as $A_p$ and $B_p$ written after Eq. (\ref{scatt_amp_2D}) of the main text, except that one has to use the $|m_l|$-dependent scattering volume $w_{1, |m_l|}$ and effective range $\alpha_{1, |m_l|}$. Then, averaging expression (\ref{sigma_2D_appendix}) over the angles $\phi_{\hat{\boldsymbol{q}}}$ and replacing $1/A_{p, |m_l|}$ with $1/A_{p, |m_l|} + i/ A'_p$, for the 2D inelastic rate constant we obtain:
 \be \label{alpha_2D_appendix}
 \begin{aligned}
 	&\alpha_{\text{2D}}(q) =\frac{16\hbar q^2}{m A'_p} \mathlarger{\sum_{|m_l| = 0,1}} \left\{ \left[ \mathlarger{\frac{1}{A'_p}} + q^2 \right]^2 \right.\\
	& \left. +  \left[ \mathlarger{\frac{1}{A_{p, |m_l|}}} + \left(B_{p, |m_l|} - \mathlarger{\frac{2 \ln l_0 q}{\pi}}\right) q^2 \right]^2 \right\}^{-1}.
\end{aligned}
\ee
Similarly to the 3D case, the above expression reproduces Eq. (\ref{alpha_2D}) of the main text if there is no $|m_l|$-dependence of the scattering parameters. In Fig. \ref{alpha2D1D_HighT_doublet}~a) we plot the thermally averaged inelastic rate constant $\left< \alpha_{\text{2D}} \right>_T$ for the quasi-2D classical gas at $T = 1$ $\mu$K as a function of magnetic field $B$. One can clearly see the emerging double peak structure of the inelastic rate constant for magnetic field oriented parallel to the plane of translational motion. Both peaks are by a factor of $2$ smaller than the single peak of the rate constant in the $|m_l|$-independent case. Positions of the peaks coincide with those found in the experiment measuring particle losses \cite{Gunter2005}.

In order to have the doubling of the resonance in the quasi-1D geometry, the magnetic field has to be neither parallel, nor perpendicular to the line of the translational motion \cite{Peng2014}. Let us consider the situation where the $B$-field forms an angle $\beta$ with the quasi-1D tube. Then the 1D scattering amplitude can be written as \cite{Peng2014} $f_{\text{1D}}(q) = -i q \left[ 1/{\cal L} + iq \right]^{-1}$ with
\be \label{Lcalligraphic}
	1/{\cal L} = \frac{ {\cal F}_0 \left[ {\cal F}_1 + {\cal G} \right] \cos^2\beta + {\cal F}_1 \left[ {\cal F}_0 + {\cal G} \right] \sin^2\beta }{{\cal F}_0 \sin^2\beta + {\cal F}_1 \cos^2\beta + {\cal G}},
\ee
where we have the functions ${\cal F}_{|m_l|} = 1/l_{p, |m_l|} + \xi_{p, |m_l|} q^2$ and ${\cal G} = {\cal D}_1/l_0 + {\cal D}_2 l_0 q^2$, with numerical constants ${\cal D}_1 \approx -0.4648$ and ${\cal D}_2 \approx 0.8316$. Here $q$ is the 1D relative momentum, and the 1D scattering parameters $l_{p, |m_l|}$ and $\xi_{p, |m_l|}$ are given by the same expressions as $l_p$ and $\xi_p$ in the main text. The only difference is that the scattering volume $w_{1, |m_l|}$ and effective range $\alpha_{1, |m_l|}$ now depend on $|m_l|$. Then, replacing $1/l_{p,|m_l|}$ with $1/l_{p,|m_l|} + i/l'_p$ and repeating the steps from section \ref{section1D}, for the 1D inelastic rate constant we obtain:
\be \label{alpha_1D_appendix}
	\alpha_{\text{1D}}(q) = \frac{8\hbar}{m} \frac{\text{Im}\left\{ 1/{\cal L} \right\} q^2}{ \left( \text{Re}\left\{ 1/{\cal L} \right\} \right)^2 + \left( \text{Im}\left\{ 1/{\cal L} \right\} + q \right)^2},
\ee
where
\begin{widetext}
\be \label{ReL}
	\text{Re}\left\{ \frac{1}{ {\cal L} } \right\} = \frac{ \left( {\cal F}_1 + {\cal G} \sin^2\beta \right)\sin^2 \beta \, \Phi^2 + \left[ 2 {\cal F}_1\left(  {\cal F}_1 + {\cal G} \right) + \left( 1/l'^2_p + {\cal G}^2 -  {\cal F}_1^2 \right)\cos^2\beta \right]\,\Phi +  {\cal F}_1 \left[ \left( {\cal F}_1 + {\cal G} \right)^2 + 1/l'^2_p  \right] }{ \sin^4\beta \, \Phi^2 + 2 \left( {\cal F}_1 + {\cal G} \right) \sin^2\beta \, \Phi + \left( {\cal F}_1 + {\cal G} \right)^2 + 1/l'^2_p},
\ee
\be	\label{ImL}
	\text{Im}\left\{ \frac{1}{ {\cal L} } \right\} = \frac{ \sin^2\beta \, \Phi^2 + 2 \left( {\cal F}_1 + {\cal G} \right) \sin^2\beta \, \Phi + \left( {\cal F}_1 + {\cal G} \right)^2 + 1/l'^2_p }{ l'_p \left[  \sin^4\beta \, \Phi^2 + 2 \left( {\cal F}_1 + {\cal G} \right) \sin^2\beta \, \Phi + \left( {\cal F}_1 + {\cal G} \right)^2 + 1/l'^2_p \right] },
\ee
\end{widetext}
with $\Phi = {\cal F}_0 - {\cal F}_1$. One can easily verify that if there is no $|m_l|$-dependence of $w_1$ and $\xi_p$, then $\Phi = 0$. Thus, we have $\text{Re}\left\{ 1/{\cal L} \right\} = 1/l_p + \xi_p q^2$ and $\text{Im}\left\{ 1/{\cal L} \right\} = 1/l'_p$ and recover expression (\ref{alpha_1D}) of the main text. Taking $\beta = 45\degree$, we plot the thermally averaged inelastic rate constant $\left< \alpha_{\text{1D}} \right>_T$ for the quasi-1D classical gas at $T = 1$ $\mu$K as a function of magnetic field $B$ in Fig. \ref{alpha2D1D_HighT_doublet}~b). We again see that the rate constant has a characteristic doublet structure. However, unlike in 3D and 2D, both peaks of $\left< \alpha_{\text{1D}} \right>_T$ are now slightly higher than the single peak of the rate constant in the $m_l$-independent case. The origin of this enhancement becomes more clear if we simplify expression (\ref{alpha_1D_appendix}) for the inelastic rate constant by omitting the terms $1/l'^2_p$ and ${\cal G}$ in Eqs. (\ref{ReL}) and (\ref{ImL}). Then, the rate constant can be written as
\begin{widetext}
\be \label{alpha_1D_appendix_simplified}
	\alpha_{\text{1D}}(q) = \frac{8 \hbar q^2}{m l'_p} \left\{ \frac{ \cos^2\beta }{ {\cal F}_0^2 + \left[ \cos^2\beta +  \mathlarger{\frac{ {\cal F}_0 }{ {\cal F}_1 } } \sin^2\beta \right]^2 q^2 } + \frac{ \sin^2\beta }{ {\cal F}_1^2 + \left[ \sin^2\beta + \mathlarger{ \frac{ {\cal F}_1}{ {\cal F}_0 } } \cos^2\beta \right]^2 q^2 } \right\}.
\ee
\end{widetext}
The term $1/l'^2_p$ is negligibly small, and by neglecting ${\cal G}$ we slightly shift positions of the peaks, since this term essentially renormalizes ${\cal F}_{|m_l|}$ (see Eq. (\ref{Lcalligraphic})). However, the behavior of the rate constant becomes much more transparent. Indeed, in Eq. (\ref{alpha_1D_appendix_simplified}) the first term corresponds to the peak for $m_l = 0$ and the second term to the peak for $|m_l| = 1$. Then, close to the resonance for $m_l = 0$ on its negative side we have ${\cal F}_0 \approx 0$, and the second term in Eq. (\ref{alpha_1D_appendix_simplified}) vanishes. The first term behaves as the rate constant given by Eq. (\ref{alpha_1D}) in the main text (where we can omit the term $1/l'_p$ in the denominator), with an extra factor of $1/\cos^2\beta$. Therefore, for $\beta = 45\degree$ the peak value corresponding to $m_l = 0$ becomes approximately by a factor of 2 larger than the single peak in the $m_l$-independent case. The resonance corresponding to $|m_l| = 1$ can be analyzed in the same way.

\begin{widetext}

\begin{figure}[H]
\subfloat{%
  \includegraphics[width=.49\linewidth]{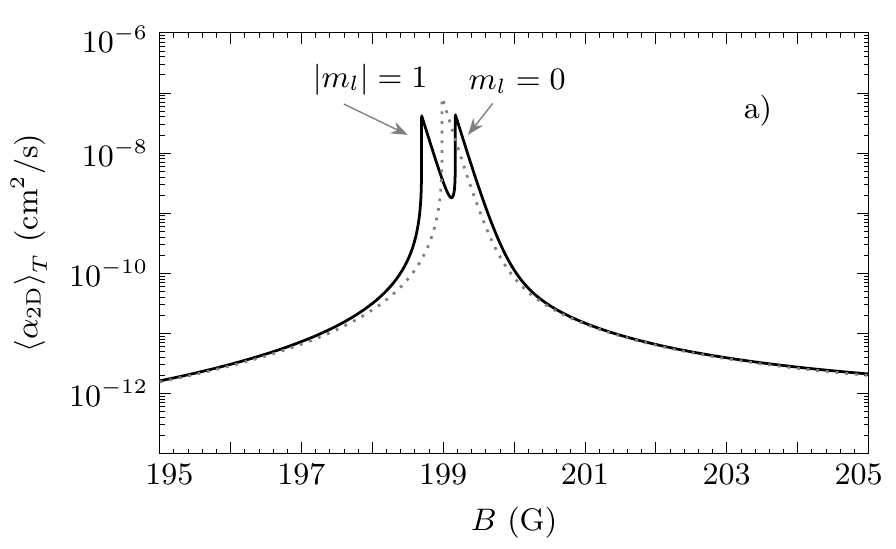}%
}\hfill
\subfloat{%
  \includegraphics[width=.49\linewidth]{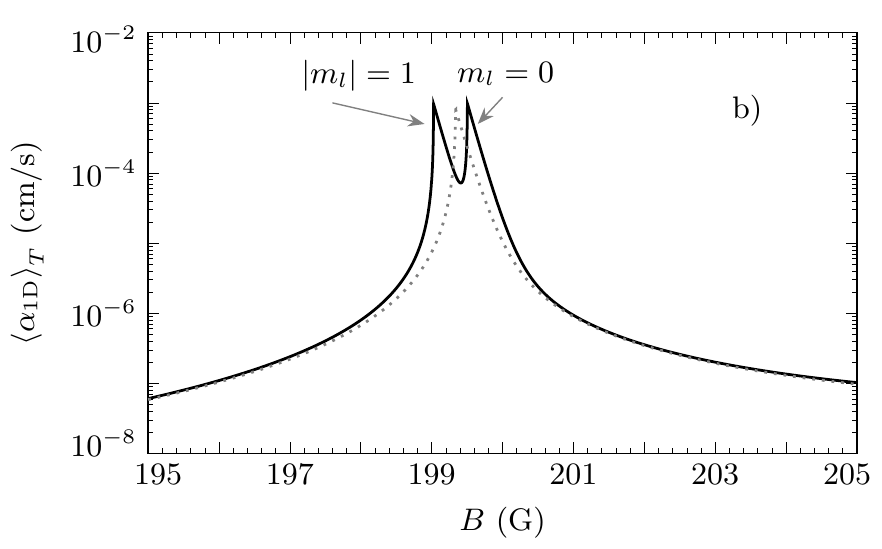}%
}
\caption{Two-dimensional inelastic rate constant $\left<\alpha_{\text{2D}}\right>_T$ in a) and one-dimensional inelastic rate constant $\left<\alpha_{\text{1D}}\right>_T$ in b) for $^{40}$K atoms in the $\ket{9/2,-7/2}$ state versus magnetic field $B$ for $T = 1$ $\mu$K and confining frequency $\omega_0 = 120$ kHz. The magnetic field is parallel to the plane of the translational motion in a) and forms an angle of $45\degree$ with the line of the translational motion in b). For comparison, grey dotted lines show the corresponding inelastic rate constants from Fig. \ref{alpha2D_HighT} b) and Fig. \ref{alpha1D_HighT} b) of the main text.}
\label{alpha2D1D_HighT_doublet}
\end{figure}

\begin{figure}[H]
\subfloat{
  \includegraphics[width=.49\linewidth]{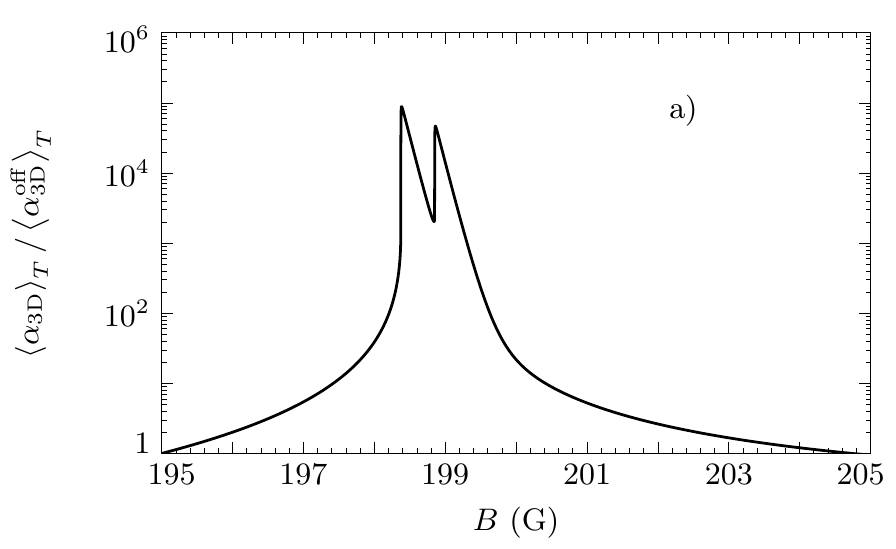}
}\hfill
\subfloat{
  \includegraphics[width=.49\linewidth]{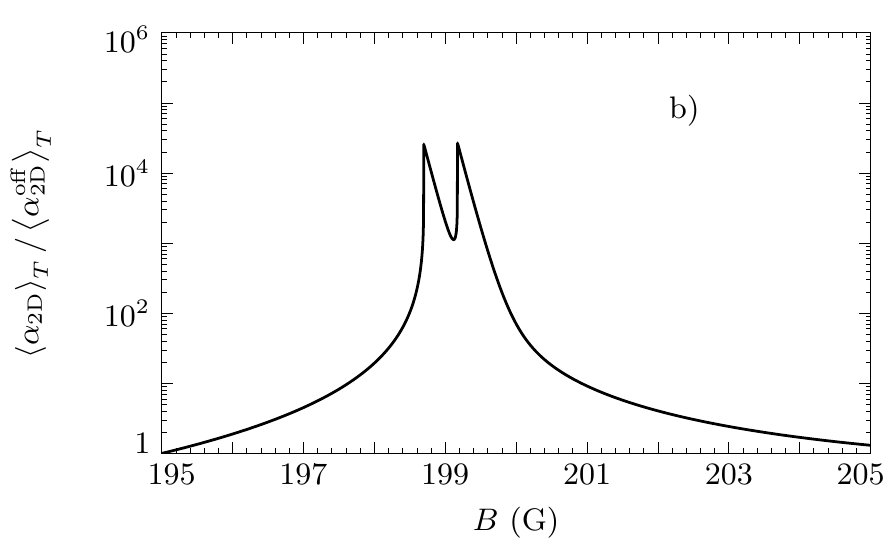}
}\hfill
\subfloat{
  \includegraphics[width=.49\linewidth]{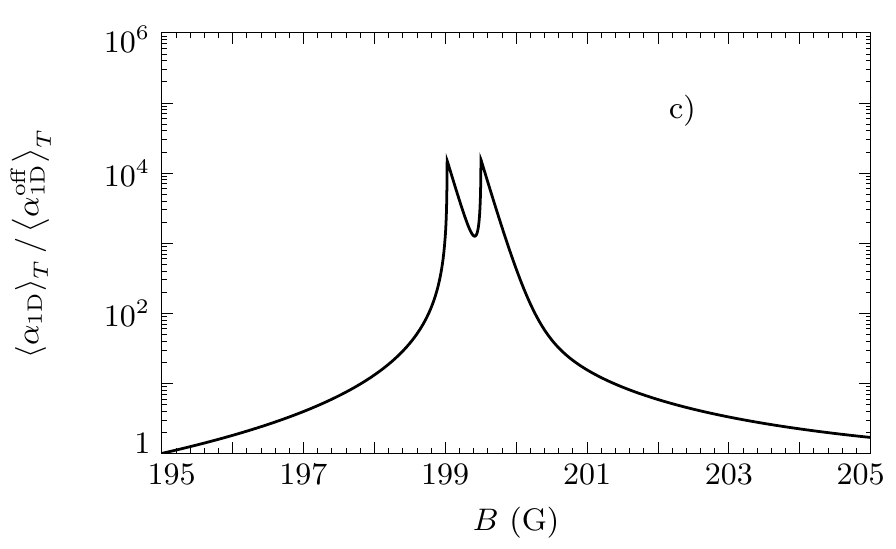}
}\hfill
\subfloat{
  \includegraphics[width=.49\linewidth]{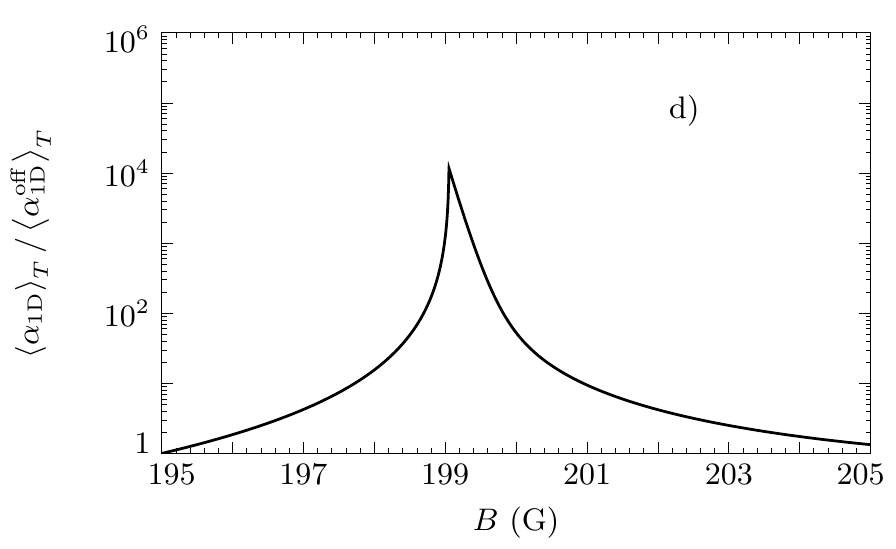}
}
\caption{Inelastic rate constant divided by its off-resonant value at a fixed filed of 195 G for $^{40}$K atoms in the $\ket{9/2,-7/2}$ state versus magnetic field $B$ for $T = 1$ $\mu$K. a) Three-dimensional case; b) Two-dimensional case with the magnetic field parallel to the plane of the translational motion; c) One-dimensional case with the magnetic field forming an angle of $45\degree$ with the line of the translational motion; d) One-dimensional case with the magnetic field perpendicular to the line of the translational motion. In Figs. b) - d) the confining frequency is $\omega_0 = 120$ kHz.}
\label{alpha_over_off_res_val}
\end{figure}

\end{widetext}

Finally, we show that the suppressed enhancement of the inelastic rate near the resonance in reduced dimensionalities is still present even if the scattering parameters are $m_l$-dependent. This is illustrated in Fig. \ref{alpha_over_off_res_val} for $^{40}$K atoms in the $\ket{9/2,-7/2}$ state at $T = 1$ $\mu$K. In Fig. \ref{alpha_over_off_res_val}~a) we plot the ratio of the 3D rate constant to its off-resonant value, $\left< \alpha_{\text{3D}} \right>_T / \left< \alpha_{\text{3D}}^{\text{off}} \right>_T$ versus magnetic field $B$. The off-resonant value is taken at a fixed field value of $195$ G. Fig. \ref{alpha_over_off_res_val} b) displays the corresponding quantity in the quasi-2D geometry with the magnetic field parallel to the plane of the translational motion. In the quasi-1D geometry with the magnetic field forming an angle of $45\degree$ with the line of the translational motion the ratio $\left< \alpha_{\text{1D}} \right>_T / \left< \alpha_{\text{1D}}^{\text{off}} \right>_T$ is plotted in Fig. \ref{alpha_over_off_res_val} c) and for the case of magnetic field perpendicular to the line of the translational motion in Fig. \ref{alpha_over_off_res_val} d). One can see the suppressed enhancement of the inelastic rate in 2D and 1D compared to 3D. This suppression is especially pronounced for the case in Fig.  \ref{alpha_over_off_res_val} d).


\begin{thebibliography}{99}

\bibitem{Gurarie2007}
	V. Gurarie and L. Radzihovsky, {\href{http://dx.doi.org/10.1016/j.aop.2006.10.009}{Ann. Phys. (Amsterdam) {\bf 322}, 2 (2007)}}.
\bibitem{Nayak2008}
	C. Nayak, S.H. Simon, A. Stern, M. Freedman, and S. Das Sarma, {\href{http://dx.doi.org/10.1103/RevModPhys.80.1083}{Rev. Mod. Phys. {\bf 80}, 1083 (2008)}}.
\bibitem{Stern2008}
	A. Stern, {\href{http://dx.doi.org/10.1016/j.aop.2005.03.001}{Ann. Phys. {\bf 323}, 204 (2008)}}.
\bibitem{Sanner2012}
	C. Sanner, E.\,J. Su, W. Huang, A. Keshet, J. Gillen, and W. Ketterle, \href{http://dx.doi.org/10.1103/PhysRevLett.108.240404}{Phys. Rev. Lett. {\bf 108}, 240404 (2012)}.
\bibitem{Pekker2011}
	D. Pekker, M. Babadi, R. Sensarma, N. Zinner, L. Pollet, M.\,W. Zwierlein, and E. Demler, \href{http://dx.doi.org/10.1103/PhysRevLett.106.050402}{Phys. Rev. Lett. {\bf 106}, 050402 (2011)}.
\bibitem{Jiang2016}
	Y. Jiang, D.\,V. Kurlov, X.-W. Guan, F. Schreck, and G.\,V. Shlyapnikov, \href{https://doi.org/10.1103/PhysRevA.94.011601}{Phys. Rev. A {\bf 94}, 011601(R) (2016)}.
\bibitem{Yang2016}
	L. Yang, X-W. Guan, and X. Cui, \href{https://doi.org/10.1103/PhysRevA.93.051605}{Phys. Rev. A {\bf 93}, 051605(R) (2016)}.
\bibitem{Regal2003}
	C.\,A. Regal, C. Ticknor, J.\,L. Bohn, and D.\,S. Jin, \href{http://dx.doi.org/10.1103/PhysRevLett.90.053201}{Phys. Rev. Lett. {\bf90}, 053201 (2003)}.
\bibitem{Ticknor2004}
	C. Ticknor, C.\,A. Regal, D.\,S. Jin, and J.\,L. Bohn, \href{http://dx.doi.org/10.1103/PhysRevA.69.042712}{Phys. Rev. A {\bf 69}, 042712 (2004)}.
\bibitem{Chevy2005}
	F. Chevy, E.\,G.\,M. van Kempen, T. Bourdel, J. Zhang, L. Khaykovich, M. Teichmann, L. Tarruell, S.\,J.\,J.\,M.\,F. Kokkelmans, and C. Salomon, \href{http://dx.doi.org/10.1103/PhysRevA.71.062710}{Phys. Rev. A {\bf 71}, 062710 (2005)}.
\bibitem{Gaebler2007}
	J.\,P. Gaebler, J.\,T. Stewart, J.\,L. Bohn, and D.\,S. Jin, \href{http://dx.doi.org/10.1103/PhysRevLett.98.200403}{Phys. Rev. Lett. {\bf 98}, 200403 (2007)}.
\bibitem{Levinsen2008}
	J. Levinsen, N.R. Cooper, and V. Gurarie, {\href{http://dx.doi.org/10.1103/PhysRevA.78.063616}{Phys. Rev. A {\bf 78}, 063616 (2008)}}.
\bibitem{JonaLasinio2008}
	M. Jona-Lasinio, L. Pricoupenko, and Y. Castin, {\href{http://dx.doi.org/10.1103/PhysRevA.77.043611}{Phys. Rev. A {\bf 77}, 043611 (2008)}}.
\bibitem{Moritz2005}
	H. Moritz, T. St\"oferle, K. G\"unter, M. K\"ohl, and T. Esslinger, \href{https://doi.org/10.1103/PhysRevLett.94.210401}{Phys. Rev. Lett. {\bf 94}, 210401 (2005)}.
\bibitem{Gunter2005}
	K. G\"unter T. St\"oferle, H. Moritz, M. K\"ohl, and T. Esslinger, \href{http://dx.doi.org/10.1103/PhysRevLett.95.230401}{Phys. Rev. Lett. {\bf 95}, 230401 (2005)}.
\bibitem{Fuchs2008}
	J. Fuchs, C. Ticknor, P. Dyke, G. Veeravalli, E. Kuhnle, W. Rowlands, P. Hannaford, and C.\,J. Vale, \href{https://doi.org/10.1103/PhysRevA.77.053616}{Phys. Rev. A {\bf 77}, 053616 (2008)}.
\bibitem{Dyke2011}
	P. Dyke, E.\,D. Kuhnle, S. Whitlock, H. Hu, M. Mark, S. Hoinka, M. Lingham, P. Hannaford, and C.\,J. Vale, \href{https://doi.org/10.1103/PhysRevLett.106.105304}{Phys. Rev. Lett. {\bf 106}, 105304 (2011)}.
\bibitem{LL}	
	L.D. Landau and E.M. Lifshitz, {\it Quantum Mechanics, Non-Relativistic Theory} (Butterworth-Heinemann, Oxford, 1999).
\bibitem{MM} N.\,F. Mott, H.\,S.\,W. Massey, {\it Theory of Atomic Collisions} (Clarendon Press, Oxford, 1965).
\bibitem{Balakrishnan1997}
	N. Balakrishnan, V. Kharchenko, R.\,C. Forrey, A. Dalgarno, \href{http://dx.doi.org/10.1016/S0009-2614(97)01052-X}{Chem. Phys. Lett. {\bf 280}, 5 (1997)}.
\bibitem{BohnPrivate}
	J. Bohn (private communication).
\bibitem{Petrov2001}
	D.\,S. Petrov and G.\,V. Shlyapnikov, \href{https://doi.org/10.1103/PhysRevA.64.012706}{Phys. Rev. A {\bf 64}, 014706 (2001)}. 
\bibitem{Pricoupenko2008}
	L. Pricoupenko, \href{http://dx.doi.org/10.1103/PhysRevLett.100.170404}{Phys. Rev. Lett. {\bf 100}, 170404 (2008)}.
\bibitem{Peng2014}
	S.-G. Peng, S. Tan, and K. Jiang, \href{https://doi.org/10.1103/PhysRevLett.112.250401}{Phys. Rev. Lett. {\bf 112}, 250401 (2014)}.
\bibitem{noteRes}
	The quantity ${\cal R}_{\text{2D}}$ remains the same if in the near-resonant regime (at $A_p<0$) we are fairly close to the 3D resonance.
\bibitem{Gao2015}
	T.-Y. Gao, S.-G. Peng, and K. Jiang, \href{https://doi.org/10.1103/PhysRevA.91.043622}{Phys. Rev. A {\bf 91}, 043622 (2015)}.
\bibitem{note3}
	In 1D one has an extra suppression of 3-body recombination of identical fermions in the off-resonant regime by a factor of $(E_F^{\text{1D}}/E_\ast)$ at $T=0$ and $(T/E_\ast)$ in the classical gas, where $E_\ast\sim 1$ mK is a characteristic molecular energy. For typical densities/temperatures this is about 3 orders of magnitude: B.D. Esry, H. Suno, and C.H. Greene, \href{http://dx.doi.org/10.1142/9789812705099_0014}{{\it The Expanding Frontier of Atomic Physics} (ICAP-2002) (World Scientific, Singapore, 2003)}; N.\,P. Mehta, B.\,D. Esry, and C.\,H. Greene, \href{http://dx.doi.org/10.1103/PhysRevA.76.022711}{Phys. Rev. A {\bf76}, 022711 (2007)}.
\end{thebibliography}
\end{document}